\documentclass[aps,prd,nofootinbib,preprintnumbers,amsmath,amssymb,latexsym,array,enumerate,letter,twocolumn,superscriptaddress]{revtex4-1}
\usepackage{hyperref}
\usepackage{graphicx}         
\usepackage{bm}               
\usepackage{amssymb}          
\usepackage{amsmath}          
\usepackage{mathtools}
\usepackage{lipsum, color}

\usepackage{amssymb,amsmath,mathrsfs,epstopdf,slashed,color}
\usepackage{tikz}
\usetikzlibrary{matrix}
\usetikzlibrary{positioning}
\usepackage{fancyhdr} 
\usepackage{lastpage} 
\usepackage{extramarks} 
\usepackage{graphicx} 
\usepackage{lipsum} 
\usepackage{amsmath}
\usepackage{amsfonts}
\usepackage{amssymb}
\usepackage{latexsym}
\usepackage{color}
\usepackage{xypic}
\usepackage{cancel,slashed}

\usepackage{graphicx}
\usepackage{placeins}
\usepackage{hyperref}
\usepackage{braket}
\usepackage{bm}
\usepackage{diagbox}
\usepackage{multirow}
\usepackage{newfile}

\begin{document}

\title{Axi-Higgs cosmology: Cosmic Microwave Background and cosmological tensions}

\author{Hoang Nhan Luu}%
 \email{hnluu@connect.ust.hk}
\affiliation{Department of Physics and Jockey Club Institute for Advanced Study, \\
	The Hong Kong University of Science and Technology, Hong Kong S.A.R., China}

\date{\today}

\begin{abstract}
Non-canonical cosmology with an uplifted Higgs vacuum expectation value~(Higgs-VEV) in the early universe is believed to provide the solution for existing tensions within the $\Lambda$CDM regime. We recently proposed a theoretical model called axi-Higgs to explore this framework. The axi-Higgs model features an ultralight axion with mass $m_a \sim 10^{-29}$ eV, which couples to the Higgs field such that the Higgs-VEV is driven by the axion background evolution. Phenomenologically, the axi-Higgs model is equivalent to $\Lambda$CDM+$m_e$+$\omega_a$ which is defined as $\Lambda$CDM with electron mass variation and the presence of an ultralight axion. In this paper, we perform Markov Chain Monte Carlo analyses to investigate the parameter space of axi-Higgs and compare it with $\Lambda$CDM, $\Lambda$CDM+$m_e$ where only electron mass variation is allowed, and $\Lambda$CDM+$\omega_a$ where only an ultralight axion is present. Combining observational data from Cosmic Microwave Background, Baryon Acoustic Oscillations, weak-lensing cosmic shear survey we found $H_0 = 69.4 \pm 1.3$~km/s/Mpc, which reduces the Hubble tension to approximately $2.4\sigma$. Including data from supernovae Type IA and local Hubble measurement from SH0ES we found $H_0 = 71.34 \pm 0.96$~km/s/Mpc. The presence of this Higgs-VEV-driving axion may be tested by atomic clock and quasar spectral measurements in the near future.
\end{abstract}

\maketitle


\section{Introduction}

The standard model of cosmology under the name $\Lambda$-cold-dark-matter ($\Lambda$CDM) has long become one of the cornerstones of modern cosmology. The $\Lambda$CDM model with six basic parameters successfully explains vastly different cosmological/astrophysical observations, spanning from the primordial state of Big Bang Nucleosynthesis~(BBN) to the afterglow of Cosmic Microwave Background~(CMB) and the large-scale structure~(LSS) of galaxies. Despite its accomplishments, multiple tensions between prediction and observation have emerged as experimental sensitivity gradually improves. Among them, the Hubble tension between late-time and early-time measurements of $H_0$ has caught increasing attention recently, most notably $H_0 = 74.03 \pm 1.42~{\rm km/s/Mpc}$ from SH0ES 2019~\cite{Riess:2019cxk} versus $H_0 = 67.36 \pm 0.54~{\rm km/s/Mpc}$ from Planck 2018~\cite{Aghanim:2018eyx}. This 4-6$\sigma$ discrepancy~\cite{Verde:2019ivm} arises from whether $H_0$ is directly measured from the local Hubble flow or indirectly inferred from CMB acoustic standard rulers. Many alternative models have attempted to resolve the Hubble tension by introducing exotic species of dark matter~(DM), dark energy~(DE), dark radition, which may modify pre-recombination or post-recombination physics. The representatives of the first category include Early Dark Energy~(EDE) model~\cite{Poulin:2018cxd, Niedermann:2019olb} or additional relativistic degrees of freedom~\cite{Kreisch:2019yzn, Buen-Abad:2017gxg, Vagnozzi:2021gjh} such as neutrinos. The second category includes cosmological models with DM-DE interactions~\cite{DiValentino:2017iww}, decaying DM~\cite{Vattis:2019efj}, emergent DE with parametrized equation of state~\cite{Li:2019yem}. Other models that does not belong to these two classes includes modified gravity~\cite{DAgostino:2020dhv}, inflationary~\cite{Tram:2016rcw}, primordial magnetic field~\cite{Jedamzik:2020krr} and many more, see~\cite{DiValentino:2021izs, Perivolaropoulos:2021jda} and references therein for an overview of the landscape of the Hubble tension solutions. Although most of these solutions could help improve $H_0$, they are sensitive to other cosmological tensions, such as the $S_8$ tension~\cite{DiValentino:2020vvd} between the CMB-inferred value of $S_8$~\cite{Aghanim:2018eyx} and the one measured by low-redshift weak-lensing experiments~\cite{Heymans:2020gsg}. Despite being statistically insignificant within $\Lambda$CDM regime~\cite{Nunes:2021ipq}, this 2-3$\sigma$ discrepancy may worsen in a few aforementioned models, see~\cite{Hill:2020osr} for an example with EDE.

Recently, we proposed the so-called axi-Higgs model to alleviate the $H_0$ and $S_8$ tensions as well as the tension of Li$^7$ abundance in BBN and the isotropic cosmic birefringence~\cite{Fung:2021wbz}. The axi-Higgs model extends the standard model of electroweak interactions to include an ultralight axion field with mass $\sim 10^{-29}$~eV. The axion field lifts the Higgs vacuum expectation value (Higgs-VEV) during the cosmological recombination time, and dynamically relaxes it to the base value of $246~{\rm GeV}$ at late times. Consequently, the electron mass is slightly deviated before recombination. Its present-day value is restored after recombination as the axion background density dilutes as cold dark matter. Phenomenologically, the axi-Higgs model is similar to $\Lambda$CDM with a varying electron mass and an ultralight scalar field, denoted as $\Lambda$CDM+$m_e$+$\omega_a$. The model fits into the category of modified-recombination models, which has been shown to be the most promising in terms of resolving the Hubble tension~\cite{Schoneberg:2021qvd}. In the literature, there have been several studies exploring this direction, starting with the $\Lambda$CDM+$m_e$ model~\cite{Planck:2014ylh, Hart:2017ndk, Hart:2019dxi} where the electron mas is varied in $\Lambda$CDM framework. Using CMB data alone, $H_0$ and $m_e$ are found positively degenerate in these studies, which implies that a higher value of $H_0$ is correlated with a higher value of $m_e$. The preferred value of $m_e$ and $H_0$ can then be constrained by BAO data, but it is worth noticing that BAO data fits $\Lambda$CDM+$m_e$ worse than $\Lambda$CDM. The authors of~\cite{Sekiguchi:2020teg} further extend this model by considering $\Lambda$CDM+$m_e$+$\Omega_k$ regime where the spatial curvature is also allowed to vary. The $\Lambda$CDM+$m_e$+$\Omega_k$ model simultaneously maintains the $m_e-H_0$ degeneracy and resolves the BAO fitting issue. Later, we will see that the presence of an ultralight axion in the axi-Higgs model affects the geometry of the universe in a similar manner to $\Omega_k$. Following the foundational work by~\cite{Fung:2021wbz}, we have proposed the leading-order perturbative approach (LPA) in~\cite{Fung:2021fcj} to study the constraints of cosmological parameters in the axi-Higgs model. This quantitatively-transparent approach is semi-analytic, providing an understanding of how small variations of cosmological parameters, especially the Higgs-VEV, impacts data fitting. Motivated by the lack of a formal analysis besides the LPA results, the standard Boltzmann analysis~(SBA) is conducted for the axi-Higgs model in this paper. This traditional approach allows comprehensive implementation of state-of-the-art data to determine the precise constraints of cosmological parameters. Thus, SBA provides a valuable test on the validity of LPA.

The rest of the paper is structured as follows. We begin with a review of the axion and Higgs-VEV evolution, from Sec.~\ref{Sec:ax_background} to Sec.~\ref{Sec:ax_perturbation}. We then discuss the CMB phenomenology of the axi-Higgs model in Sec.~\ref{Sec:cmb_pheno}. After introducing observational data and methodology in Sec.~\ref{Sec:data_set} and Sec.~\ref{Sec:methodology}, we present the main findings of the paper in Sec.~\ref{Sec:models_compare} where the axi-Higgs model is compared with $\Lambda$CDM, $\Lambda$CDM+$m_e$, $\Lambda$CDM+$\omega_a$. Sec.~\ref{Sec:conclusion} provides conclusion and outlook on some questions beyond the scope of this work. In App.~\ref{App:aH_compare} we compare parameter constraints for the axi-Higgs model with different axion masses while App.~\ref{App:ah_lcdmmeax} compares the axi-Higgs model with its phenomenological model $\Lambda$CDM+$m_e$+$\omega_a$. App.~\ref{App:EFA} is devoted to test the effective fluid approximation applied to the axion equations of motion. Finally, App.~\ref{App:supplementary} provides supplementary figures and tables for the main text.


\section{Axi-Higgs physics} \label{Sec:ah_physics}

Since our focus in this paper is on the physics of CMB, we shall concentrate on the cosmic epoch around the recombination era. The axi-Higgs model is characterized by an ultralight axion field $\phi$ coupled to the electroweak Higgs doublet $\Phi$, which reads~\cite{Fung:2021wbz}
\begin{align}
V(\phi)  &= m^2 \phi^2/2  + \left|m_s^2F(\phi) - \kappa \Phi^{\dagger} \Phi\right|^2 + V_a, \label{Eq:potential_Higgs} \\  
F(\phi) &= (1+\delta v)^2 = 1+ C \frac{\phi^2}{M_{\rm Pl}^2}, \label{Eq:delta_v}
\end{align}
where the parameters $m_s$ and $\kappa$ are fixed by the present-day Higgs-VEV $v_0=246$ GeV and the Higgs boson mass $m_{\Phi}=125$ GeV; $\delta v=\Delta v/v_0=(v-v_0)/v_0$ is the fractional shift of $v$ from its present-day value; $M_{\rm Pl} \simeq 2.4 \times 10^{18}$~GeV is the reduced Planck mass. After the shift symmetry is spontaneously broken by non-perturbative dynamics, the axion field effective potential is given by
	\begin{align}
		V_a(\phi) = m_a^2f_a^2 \left[ 1 - \cos \left( \dfrac{\phi}{f_a} \right) \right],
	\end{align}
	where the axion mass $m_a \sim 10^{-29}-10^{-30}~{\rm eV}$ and decay constant $f_a \sim 10^{17}~{\rm GeV}$ are typically of our interest. Since the axion is coupled to the Higgs, radiative corrections to $m_a$ might raise concern. However, we expect that these corrections vanish above the axion characteristic scale $\Lambda_a \simeq \sqrt{m_a f_a}$ as the shift symmetry is restored, or equivalently, as the axion becomes massless~\cite{Fung:2021wbz}.

	The perfect square of the Higgs potential in Eq.~\eqref{Eq:potential_Higgs} assures that its contribution to the vacuum energy density stays at exactly zero, a necessary (but not sufficient) requirement for a naturally exponentially small cosmological constant. This perfect square form is crucial in the axi-Higgs model for two reasons. (i)~Since the axion $\phi$ and the Higgs field $\Phi$ are coupled, their cosmological evolution is closely tied together. However, the impact of the Higgs field on the axion evolution is negligible. Therefore, we can treat the axion evolution as if it is decoupled from the Higgs~\cite{Fung:2021wbz}. (ii) Any higher-order corrections of the electroweak Higgs contribution to the vacuum energy are absent~\cite{Li:2020rzo}.
 
\subsection{Axion Background Dynamics} \label{Sec:ax_background}
Providing the axion evolution starts near the bottom of its potential, the Lagrangian is written as
\begin{align}
\mathcal{L} \simeq -\dfrac{1}{2} (\partial\phi)^2 - \dfrac{1}{2}m_a^2\phi^2. \label{Eq:potential}
\end{align}
In a spatially flat and homogeneous Friedmann-Lemaitre-Robertson-Walker (FLRW) universe, the axion background equation of motion reads
\begin{align}
\ddot{\phi} + 2\mathcal{H}\dot{\phi} + m_a^2a^2\phi = 0, \label{Eq:ax_eom}
\end{align}
with the corresponding energy density and pressure given by
\begin{align}
\rho_a &= \dfrac{1}{2}a^{-2}\dot{\phi}^2 + \dfrac{1}{2}m_a^2\phi^2, \\
p_a &= \dfrac{1}{2}a^{-2}\dot{\phi}^2 - \dfrac{1}{2}m_a^2\phi^2,
\end{align}
where $\dot{} \equiv d/d \eta$ denotes the derivative with respect to conformal time; the conformal Hubble is $\mathcal{H} \equiv \dot{a}/a =aH$. At the beginning, the axion field slowly rolls down the potential due to the large Hubble friction as its energy density stays constant similarly to dark energy. Until a transition moment when the scale factor of the universe satisfying
\begin{align}
m_a \simeq \xi H(a_\text{osc}), \label{Eq:ax_osc}
\end{align}
the axion field begins its damped oscillations over cosmological time scales, effectively contributes to the dark matter density. Note that, in this stage, the axion is treated with the effective formalism as in~\cite{Hlozek:2014lca, Hlozek:2017zzf} to avoid resolving rapid oscillations on time scales smaller than the Hubble time. More specifically, we approximate
\begin{align}
\dfrac{1}{2}a^{-2}\dot{\phi}^2 \simeq \dfrac{1}{2}m_a^2\phi^2, \quad \rho_a \simeq \dfrac{\rho_a(a_\text{osc}) a_\text{osc}^3}{a^3}, \quad p_a \simeq 0.
\end{align}
The transition coefficient $\xi$ is roughly of order one, which is chosen to be $\xi = 3$ as the standard value in this work. We briefly discuss other choices in App.~\ref{App:EFA}.

\subsection{Higgs-VEV on electron mass} \label{Sec:electron_mass}
Expanding Eq.~\eqref{Eq:delta_v} up to the lowest order, we see that the Higgs-VEV is driven by the axion background evolution, which deviates from its reference value by
\begin{align}
\delta v = \dfrac{C\phi^2}{2 M^2_\text{pl}}. \label{Eq:VEV-evolve}
\end{align}
When $H(z) \gg m_a$, the initial field $\phi_\text{ini}$ stays unchanged, which raises the Higgs-VEV, $v > v_0$, before and during recombination. The value of $\phi$ starts dropping when $H(z) \lesssim \xi m_a$, which drives $v$ evolution towards $v_0$ such that
\begin{align}
\delta v \simeq \delta v_\text{osc} (a_\text{osc}/a)^3,
\end{align}
since $\delta v \propto \phi^2 \propto \rho_a \propto a^{-3}$ after the transition. The time dependence of the Higgs-VEV induces a universal evolution of particle masses, electroweak coupling constants, QCD confinement scales, etc~\cite{Flambaum:2007mj, Dent:2007zu, Berengut:2009js, Walker-Loud:2014iea, Mori:2019cfo, Clara:2020efx}. Early-time physics, especially BBN and CMB, must be extensively modified to account for this unconventional change. This work, however, is devoted to investigating potential implications on CMB alone. We further omit the variation of the proton mass, which presumably makes our results unaffected because $\delta m_p / \delta v \sim \mathcal{O}(10^{-3})$. Thus, varying the Higgs-VEV after BBN era only shifts the electron mass 
\begin{align}
\delta v = \delta m_e,
\end{align}
which alters several atomic constants involved in recombination. In the effective three-level atomic model~\cite{Zeldovich:1968qba, Peebles:1968ja, Seager:1999km} those constants are: the hydrogen and helium energy levels $E_i$; the Thompson scattering cross section $\sigma_T$; the two-photon decay rate $A_{2\gamma}$, the effective Lyman-$\alpha$ transition rate $P_S A_{1\gamma}$; the baryon temperature $T_\text{eff}$; the effective recombination rate $\alpha$ and photoionization rate$\beta$ with their dependencies on the electron mass given by~\cite{scoccola2009wmap, Planck:2014ylh, Chluba:2015gta}
\begin{gather*}
E_i \propto m_e, \quad \sigma_T \propto m_e^{-2}, \quad A_{2\gamma} \propto m_e, \quad P_S A_{1\gamma} \propto m_e^3, \\
\alpha \propto m_e^{-2}, \qquad \beta \propto m_e, \quad T_\text{eff} \propto m_e^{-1}.
\end{gather*}
We note that varying the fine-structure constant yields similar effects~\cite{Kaplinghat:1998ry}, though it is not motivated by the axi-Higgs model.

With the axion mass $m_a$ fixed, the axi-Higgs model is characterized by two theoretical parameters: $\phi_\text{ini}$ and $C$, namely the initial field value and the ``coupling constant" of the Higgs with the axion. We can equivalently convert them to two phenomenological parameters which are more relevant to cosmology: the axion relic density $\omega_a$ and the initial Higgs-VEV ratio $(v/v_0)_\text{ini}$, defined by
\begin{align}
\omega_a &= F(\phi_\text{ini}) \propto \phi^2_\text{ini}, \\
\quad (v/v_0)_\text{ini} &= 1 + C\phi^2_\text{ini}/(2M^2_\text{pl}).
\end{align}
We consider the axion abundance negligible compared to the total cosmic budget at all time so that the backreaction on the Hubble flow can be approximately neglected in Eq.~\eqref{Eq:ax_eom}.

\subsection{Axion Perturbations} \label{Sec:ax_perturbation}

The presence of an axion field sources scalar perturbations, which are written in the synchronous gauge as
\begin{align}
ds^2 = a^2(\eta)\left[ -d\eta^2 + (\delta_{ij} + h_{ij})dx^i dx^j \right],
\end{align}
with the axion perturbation equations read~\cite{Hlozek:2014lca, Hlozek:2016lzm, Hlozek:2017zzf}
\begin{align}
\dot{\delta}_a &= -ku_a - (1+w_a)\dot{h}/2  \nonumber \\
&\qquad - 3\mathcal{H}(1-w_a)\delta_a - 9\mathcal{H}^2(1-c^2_\text{ad})u_a/k, \label{Eq:ax_per} \\
\dot{u}_a &= 2\mathcal{H}u_a + k\delta_a + 3\mathcal{H}(w_a - c^2_\text{ad})u_a, \label{Eq:ax_heat}
\end{align}
where $h$ is the trace of $h_{ij}$. The density contrast and heat flux are defined, respectively, as
\begin{align}
\delta_a \equiv \delta\rho_a/\rho_a, \qquad u_a \equiv (1+w_a)v_a.
\end{align}
Here, the equation of state and adiabatic sound speed are background-dependent quantities
\begin{align}
w_a \equiv \dfrac{p_a}{\rho_a}, \qquad c^2_\text{ad} \equiv \dfrac{\dot{p}_a}{\dot{\rho}_a} = 1 + \dfrac{2m_a^2a^2\phi}{3\mathcal{H}\dot{\phi}}.
\end{align}
After axion oscillations commence, Eqs.~\eqref{Eq:ax_per} and~\eqref{Eq:ax_heat} are effectively transformed to
\begin{align}
\dot{\delta}_a &= -ku_a -\dot{h}/2 - 3\mathcal{H}c^2_a \delta_a - 9\mathcal{H}^2 c^2_a u_a/k, \\
\dot{u}_a &= -\mathcal{H} u_a + k c^2_a \delta_a + 3 c^2_a \mathcal{H} u_a,
\end{align}
where the effective axion sound speed is
\begin{align}
c^2_a = \dfrac{k^2/(4m_a^2 a^2)}{1 + k^2/(4m_a^2 a^2)}.
\end{align}
The above two effective perturbation equations are derived under an ansatz that the background axion field and its perturbation can be expanded in terms of harmonic functions with a fixed frequency of order $m_a$~\cite{Hwang:2009js, Hwang:2012}
\begin{align}
\phi &= a^{-3/2} \left[ A(\eta) \cos(m_a\eta) + B(\eta) \sin(m_a\eta) \right], \\
\delta \phi &= C_k(\eta) \cos(m_a\eta) + D_k(\eta) \sin(m_a\eta),
\end{align}
satisfying the oscillation-averaged condition $A D_k = B C_k$. Additionally, the evolution of metric perturbations is considered on time scales much longer than the intrinsic axion Compton period, i.e. $\eta \gg m_a^{-1}$. For simplicity, we focus solely on adiabatic initial conditions, i.e.
\begin{align}
\delta_a(\eta_\text{ini}) = 0, \qquad u_a(\eta_\text{ini}) = 0,
\end{align}
so the isocurvature mode of the primordial axion scalar perturbations produced during inflation is assumed to be negligible.

\subsection{Axi-Higgs cosmological implications} \label{Sec:cmb_pheno}

The Higgs-VEV-driving axion affects various cosmological observables from early to late times, which leaves distinctive phenomenology on the CMB spectra and matter power spectrum of the axi-Higgs model.

Firstly, the electron mass variation alters the standard recombination history. Heavier electrons speed up hydrogen and helium recombination, broadening the width of the photon visibility function and shifting its peak to higher redshifts. Recombination is triggered earlier and lasts longer. The sound horizon is also shorten as a consequence.

Secondly, the ultralight axion influences the cosmic budget at both background and perturbation level, see Fig.~\ref{Fig:aH_efa}. Due to DE-like nature, the $10^{-29}$-eV axion leaves minor impacts on the background expansion before recombination era when the universe was dominated by radiation and matter. After the onset of oscillations, the axion background density dilutes as CDM, therefore it mimics the CDM behavior on the late-time Hubble flow. On the other hand, axion perturbations start evolving even at early times. The horizon-sized modes always grow while the sub-horizon modes whose wavelengths are shorter than the Jeans scale are suppressed initially. The axion mode evolution follows the same redshift scaling as CDM at late times.

\begin{figure*}[!ht]
	\centering
	\includegraphics[scale=0.5]{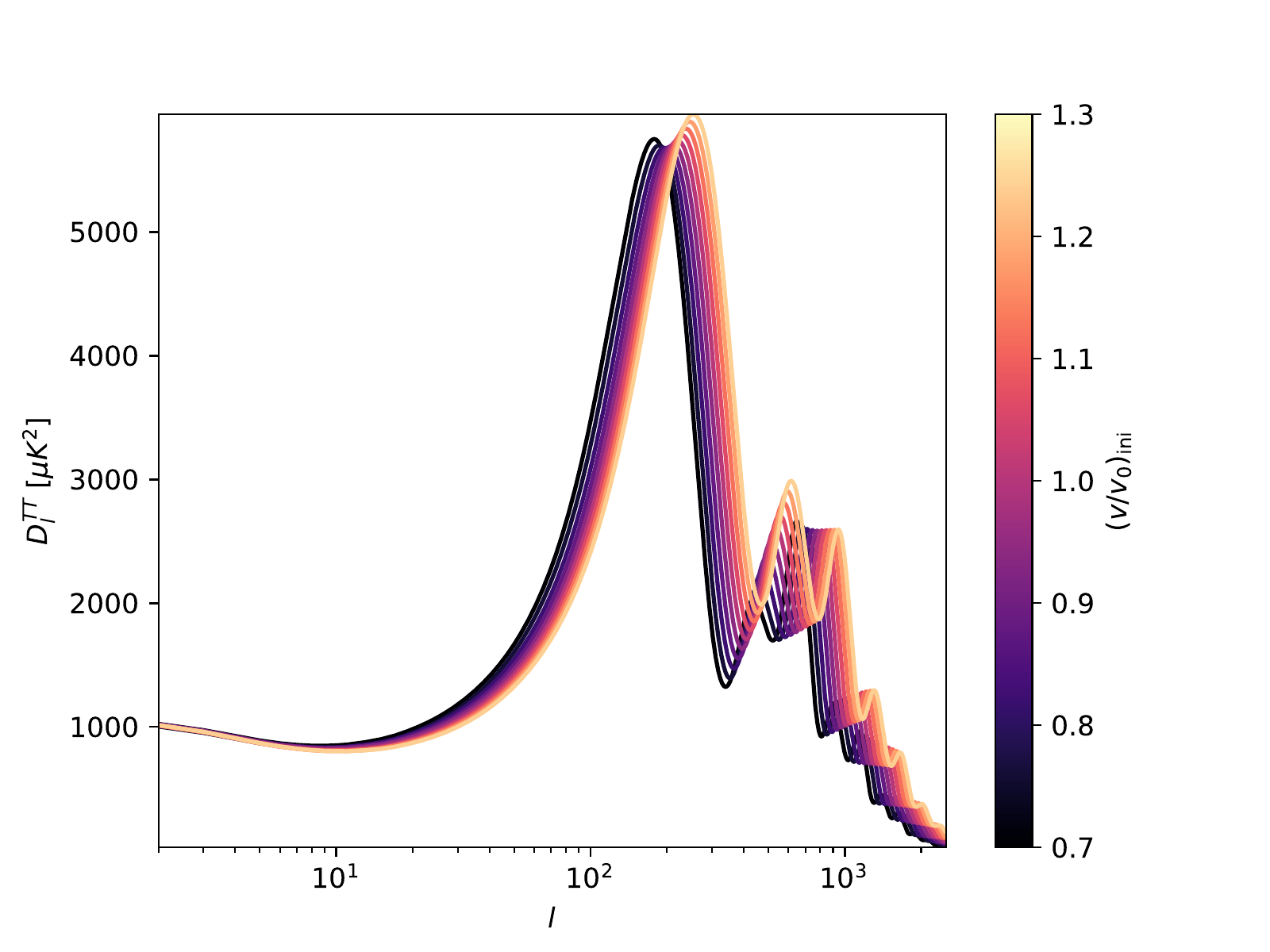}\includegraphics[scale=0.5]{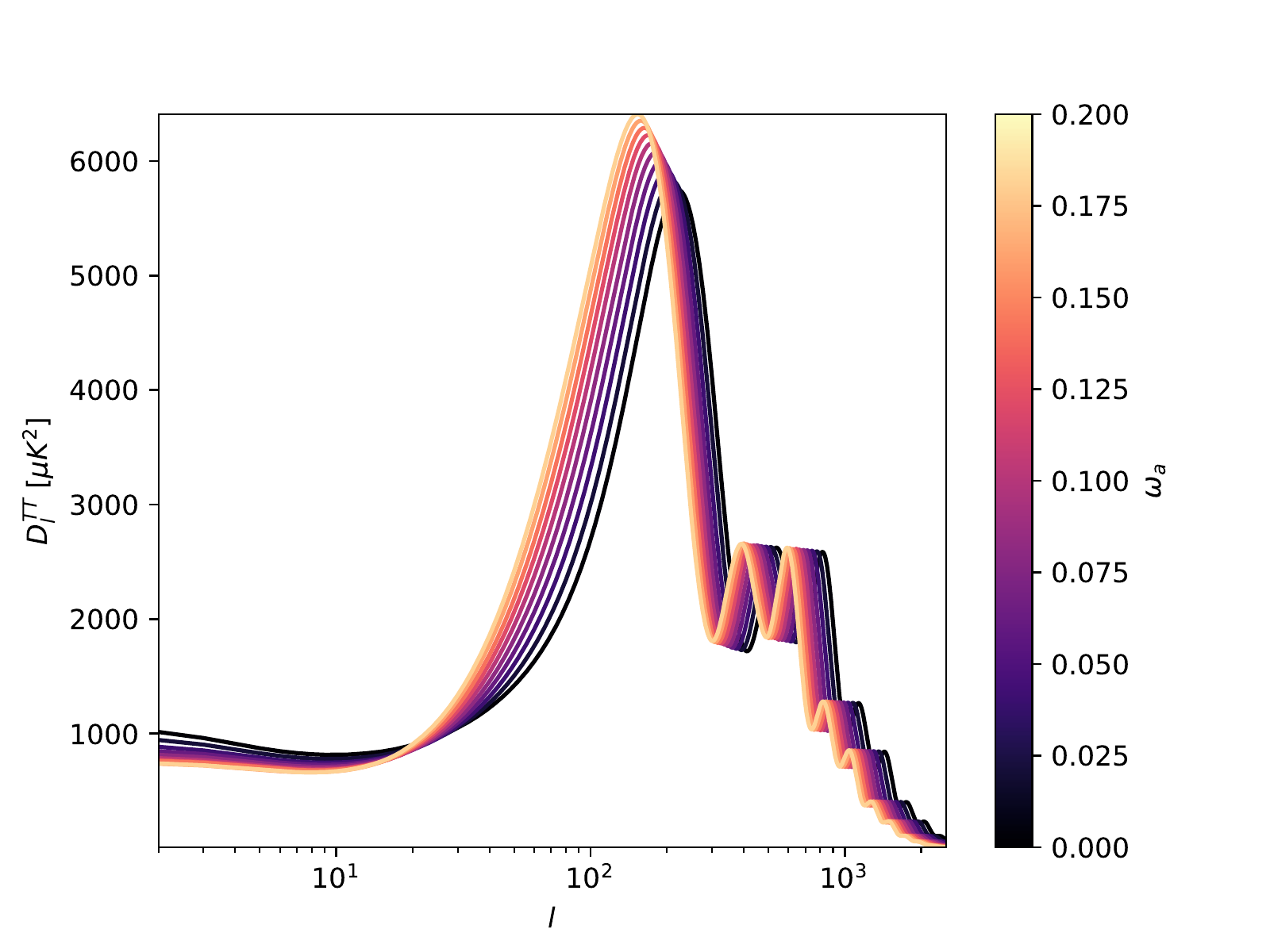}
	\includegraphics[scale=0.5]{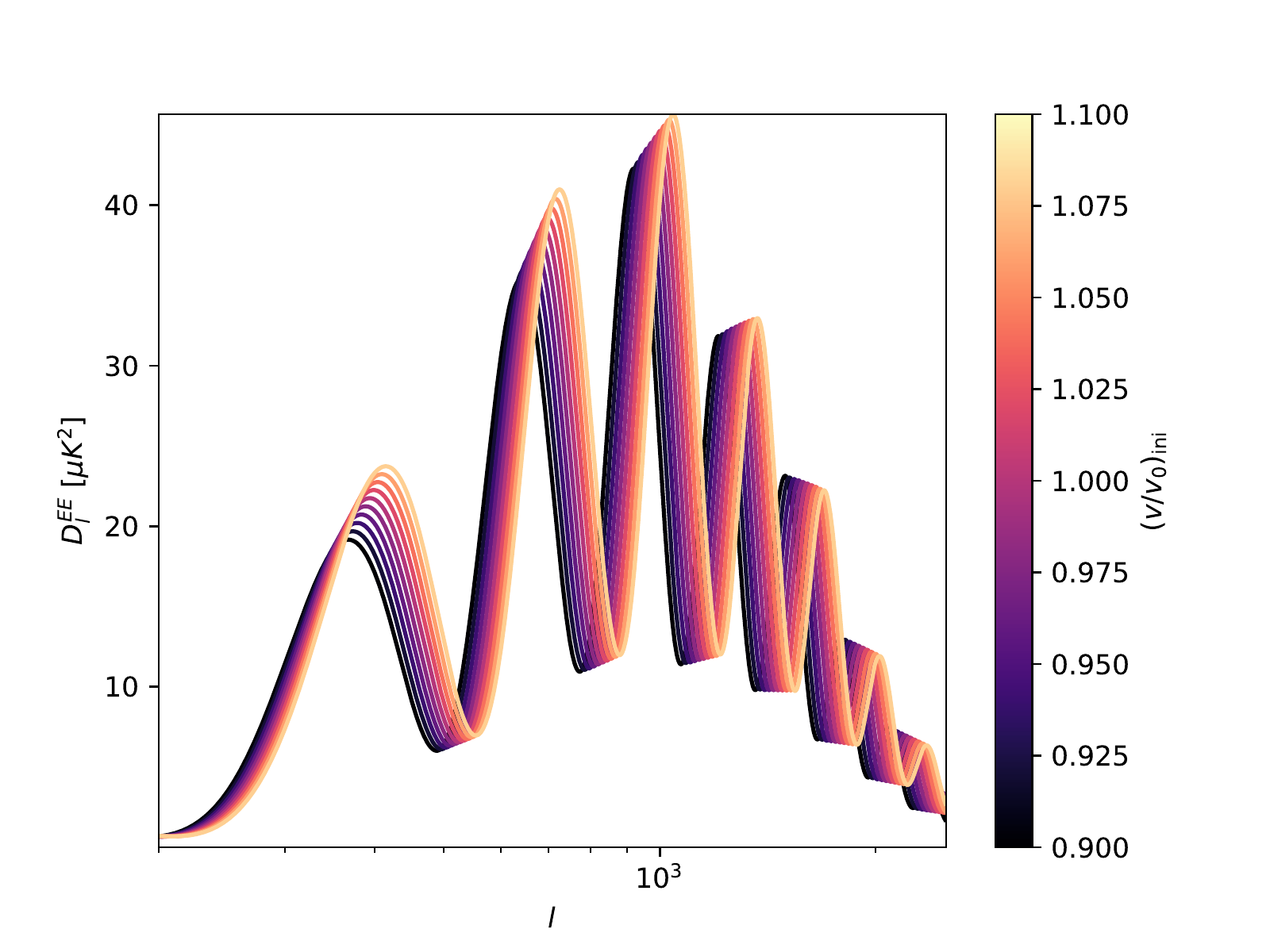}\includegraphics[scale=0.5]{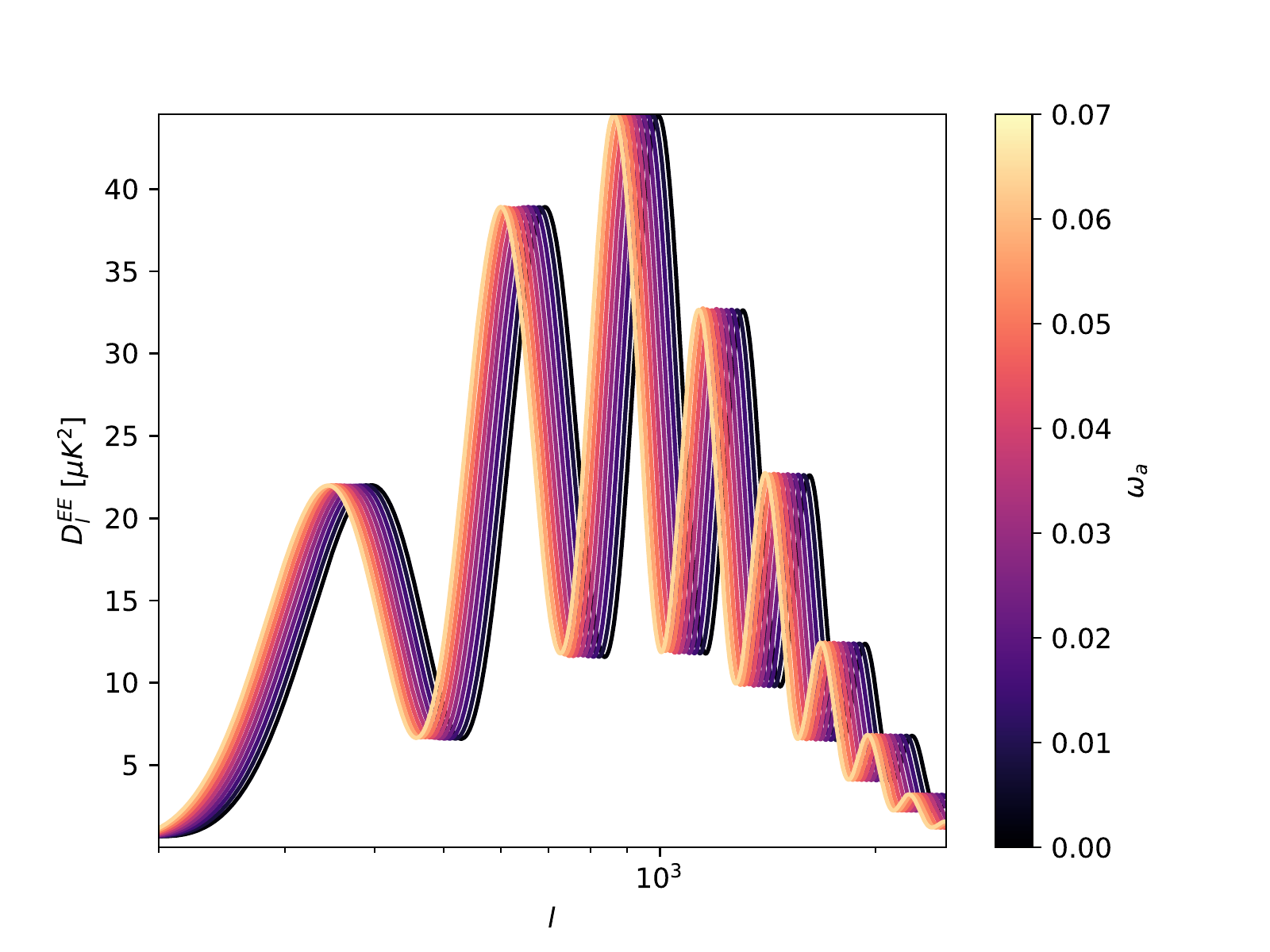}
	\includegraphics[scale=0.5]{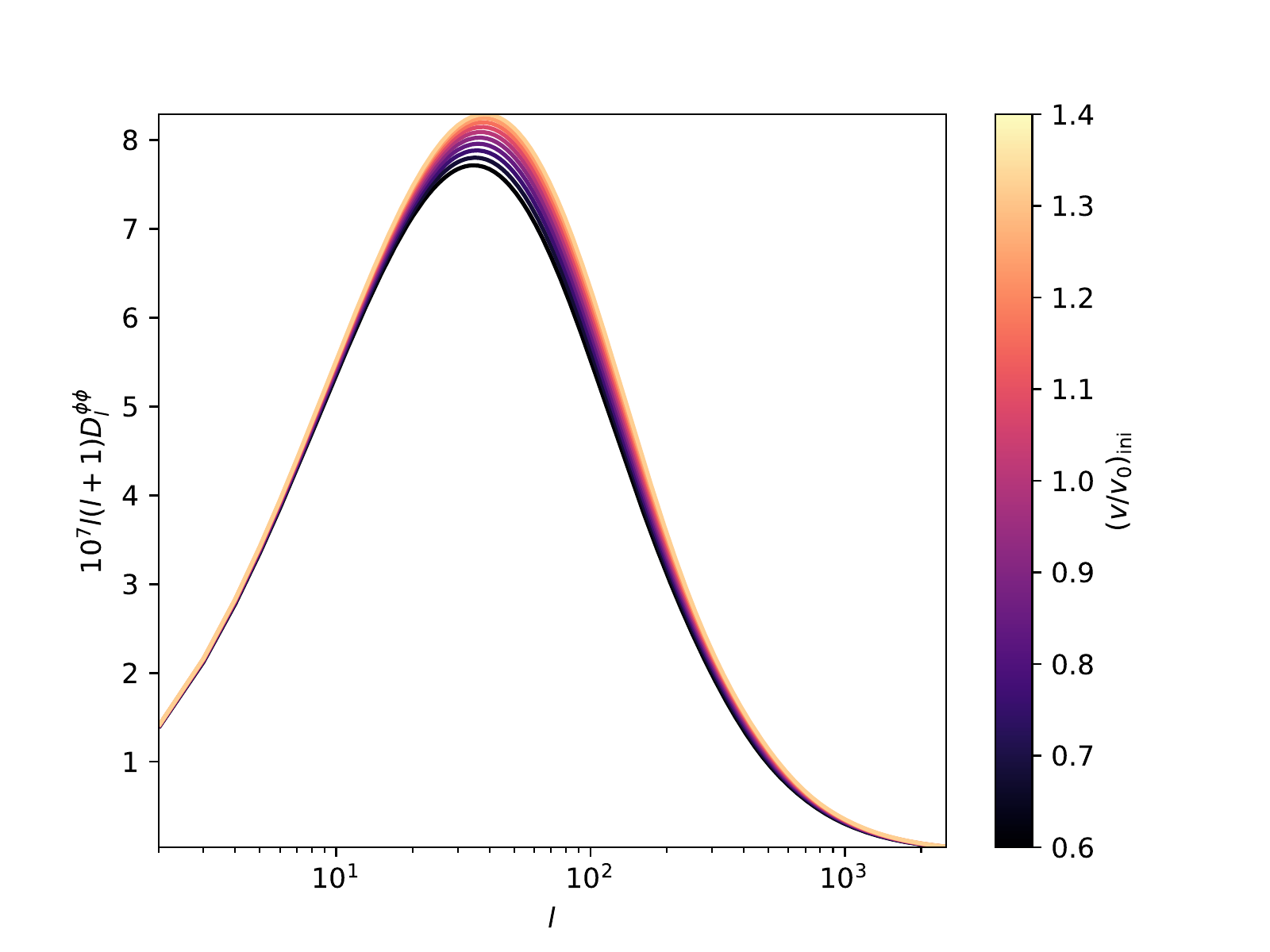}\includegraphics[scale=0.5]{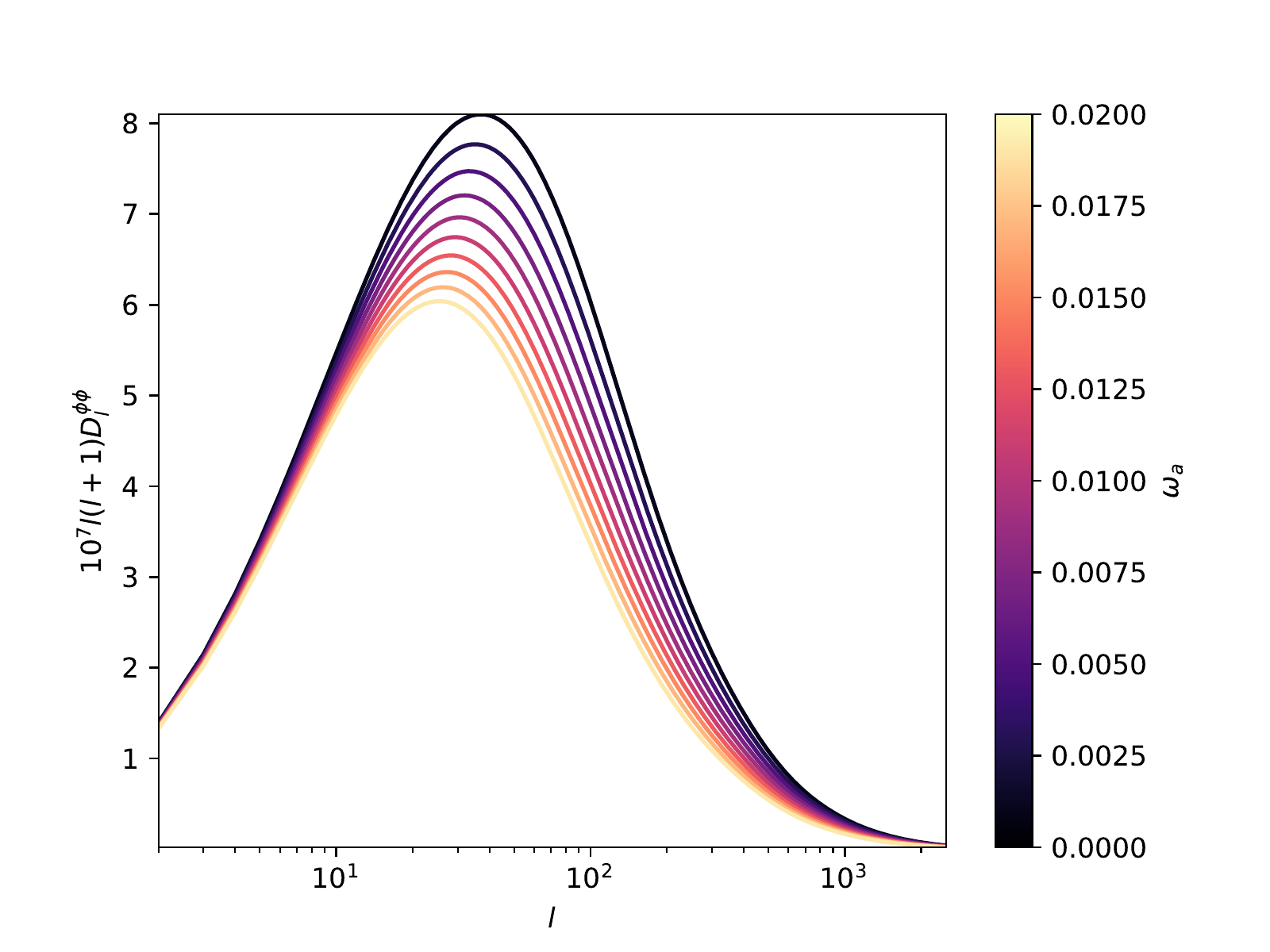}
	\caption{From top to bottom panels: the TT, EE and $\phi\phi$ power spectra in the axi-Higgs model, generated by {\tt aHCAMB}. The rescaled spectrum is related to the original one by $D_l \equiv l(l+1)C_l / 2\pi$. The color bars illustrate the range of $(v/v_0)_{\rm ini}$~\textit{(left)} and $\omega_a$~\textit{(right)} over which spectra variations are shown in each panel.}
	\label{Fig:cmb_spectra}
\end{figure*}

\begin{figure*}[!ht]
	\centering
	\includegraphics[scale=0.53]{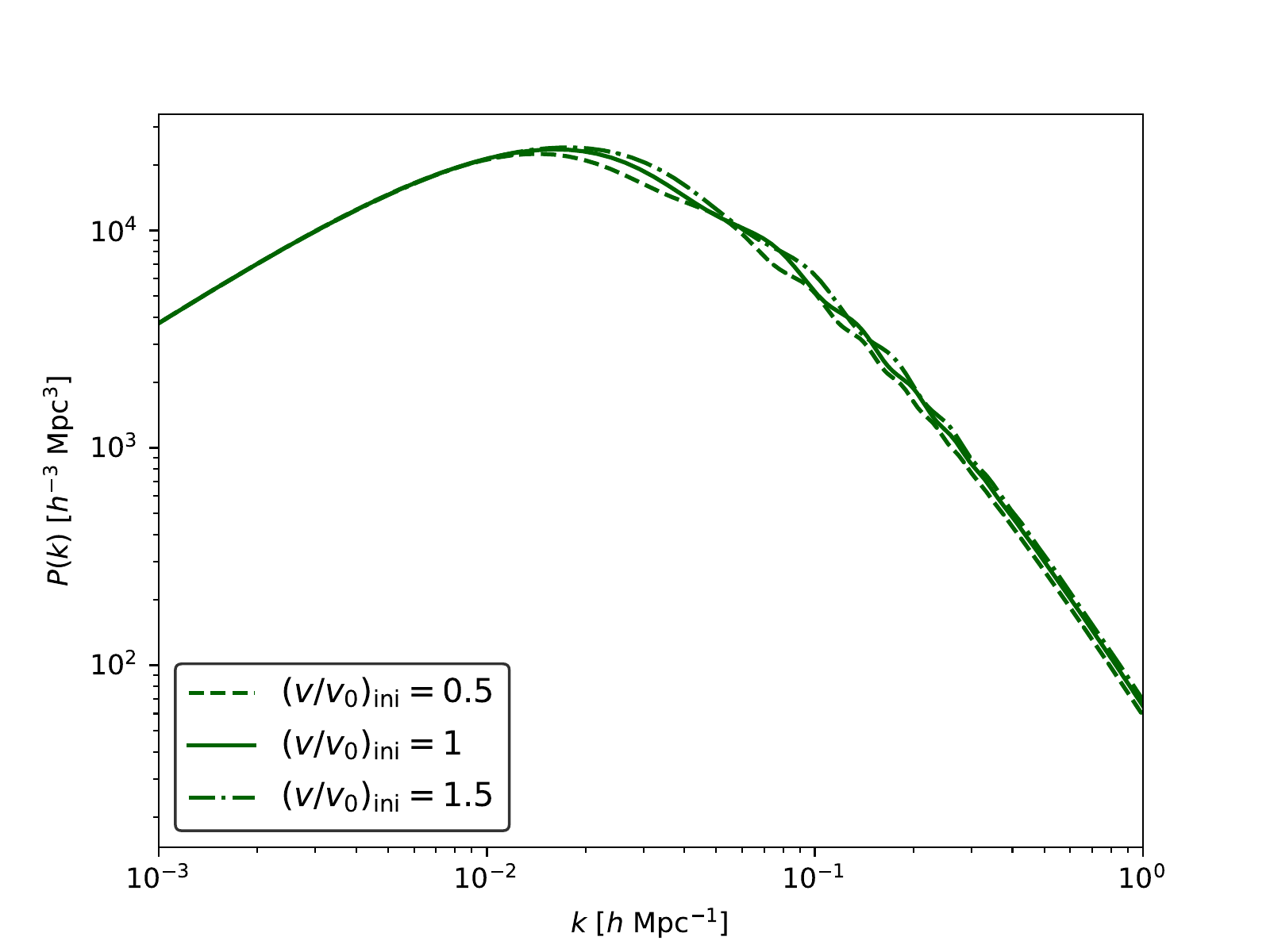}\includegraphics[scale=0.53]{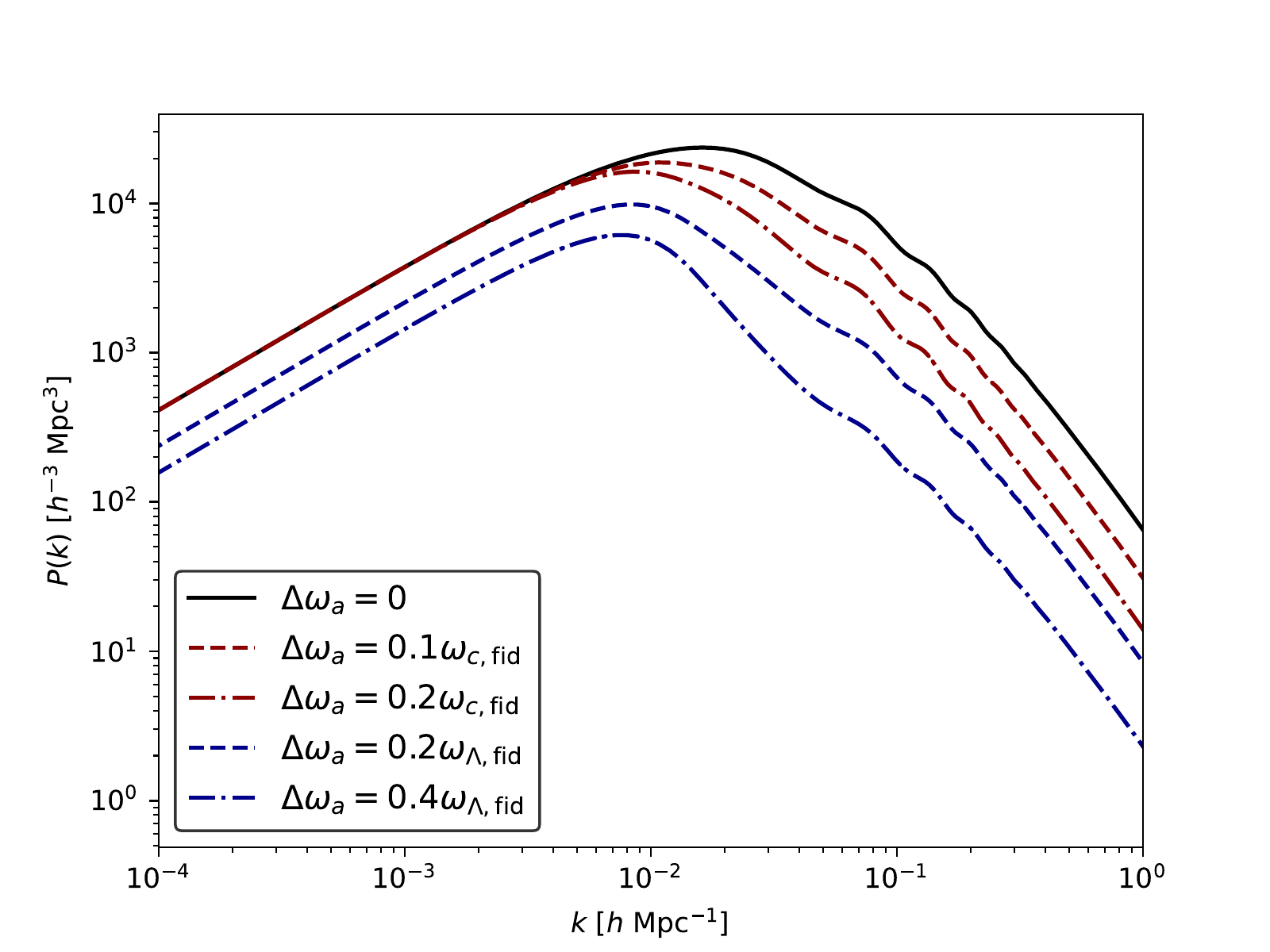}
	\caption{The matter power spectrum observed at $z=0$ in the axi-Higgs model. (\textit{Left})~The curves are plotted with three different values of $(v/v_0)_{\rm ini}$, which are almost indistinguishable from each other. (\textit{Right})~The curves are plotted with $\omega_a = \omega_{a,{\rm fid}} + \Delta \omega_a$. The red ones and the blue ones have $\omega_c = \omega_{c, {\rm fid}} - \Delta \omega_a$ and $\omega_\Lambda = \omega_{\Lambda, {\rm fid}} - \Delta \omega_a$, respectively. Other parameters are fixed to their fiducial values.}
	\label{Fig:matter_spectrum}
\end{figure*}

Fig.~\ref{Fig:cmb_spectra} and Fig.~\ref{Fig:matter_spectrum} show the CMB auto-correlation function of temperature, polarization, lensing deflection and the matter power spectrum with respect to $(v/v_0)_{\rm ini}$ and $\omega_a$. Except for the varying parameter, input parameters are fixed to a set of fiducial values, which are chosen as follows
\begin{gather}
\omega_b = 0.02238, \quad \omega_c = 0.1201, \quad H_0 = 67.32, \\
10^9A_s = 2.1, \quad n_s = 0.9660, \quad \tau_{\rm reio} = 0.0543, \\
(v/v_0)_{\rm ini} = 1.01, \quad \omega_a = 0.001, \quad m_a = 10^{-29}~{\rm eV},
\end{gather}
where the dimension [km/s/Mpc] of the Hubble constant $H_0$ is implicitly assumed hereafter. \\

From the top panels of Fig.~\ref{Fig:cmb_spectra}, we observe that the acoustic peaks of CMB spectra shift to high-l values as the initial Higgs-VEV $(v/v_0)_{\rm ini}$ is uplifted. Meanwhile, increasing $\omega_a$ drags them back to the reverse direction, which implies that the original peak positions may retain once these two parameters are turned on simultaneously. However, the peak heights are also dragged up, which requires variations of other parameters to keep the whole spectrum unchanged, e.g. increasing $\omega_c$ can suppress the peaks, see Fig.~\ref{Fig:cmb_spectra_extra}. We also notice that $\omega_a$ and $H_0$ share similarity in terms of spectrum variations, see Fig.~\ref{Fig:cmb_spectra_extra}, most notably the fact that increasing $\omega_a/H_0$ pushes the acoustic peaks toward larger angular scales. These two parameters are, therefore, oppositely degenerate.

In the middle panels of Fig.~\ref{Fig:cmb_spectra}, we see that the EE spectra are more sensitive to $(v/v_0)_{\rm ini}$ and $\omega_a$ variations compared to the TT spectra as the same order-of-magnitude deviations can be seen over smaller ranges of both parameters.

The bottom panels of Fig.~\ref{Fig:cmb_spectra} shows that the lensing power spectrum is stable with non-standard $(v/v_0)_{\rm ini}$ values but extremely suppressed with $\omega_a$. This suppression is caused by smaller matter perturbation amplitudes, which can be originated from the following scenarios. (i) Adding $\omega_a$ decreases the dark energy density $\omega_\Lambda$ given a fixed value of $H_0$. The universe with a lower amount of DE appears to have a shortened age, which is not old enough for matter clustering to form large-scale structure. As a result, the matter power spectrum is overall suppressed on every scales. (ii) If we instead replace a fraction of CDM by axion to keep DE fixed, the suppression of $C^{\phi\phi}_l$ is similar as the matter power spectrum is only suppressed on scales smaller than the Jeans scale~\cite{Hui:2016ltb, Marsh:2015xka}. These features are illustrated in the right panel of Fig.~\ref{Fig:matter_spectrum}. Eventually, we would observe CMB photons propagating from the last scattering surface less likely to be deflected by the intervening structure before reaching the Earth. Therefore, a substantial amount of axion relic density with $m_a \sim 10^{-29}$ eV is strictly prohibited by the current CMB lensing data.

\section{Standard Boltzmann Analysis} \label{Sec:sba}

We modify the standard Boltzmann solver {\tt CAMB}~\cite{Lewis:1999bs, Howlett:2012mh} to account for many aforementioned features of the axi-Higgs model. We name this numerical implementation {\tt aHCAMB} and make it publicly available ~\footnote{https://github.com/lhnhan/aHCAMB.git}.

More specifically, we incorporate an ultralight axion field at background expansion and linear perturbations with adiabatic initial conditions. The effective fluid approximation is applied for the axion after its transition redshift $a_{\rm osc}$, similarly to {\tt axionCAMB}~\cite{Hlozek:2014lca}. The axion mass is bounded to $10^{-26}~{\rm eV} < m_a < 10^{-32}~{\rm eV}$, so that axion perturbations do not significantly contribute to the matter power spectrum but still source metric perturbations. The axion background field is treated as matter in calculating relic density, i.e. $\omega_m = \omega_b + \omega_c + \omega_\nu + \omega_a$. We use the shooting method to obtain the $\phi_\text{ini}$ given $\omega_a$ as inputs, with a tolerance of $\Delta \omega_a \simeq 10^{-5}$ for numerical efficiency, so that the code will treat $\omega_a < 10^{-5}$ as $\omega_a = 0$.
	
Furthermore, we compute the free electron fraction $x_e \equiv n_e/n_{H,0}$ from {\tt Recfast++}~\cite{Seager:1999bc, Chluba:2010ca, Rubi_o_Mart_n_2010, Chluba_2010_a, Chluba_2010_b}, which is modified to include the time dependence of the electron mass following Eq.~\ref{Eq:VEV-evolve}, and the axion contribution to the Hubble function. Additionally, we scale the Thompson scattering cross-section, the Compton cooling term in the baryon-photon coupled equations and the reionization optical depth with respect to $m_e$ (or $v/v_0$) accordingly.

\subsection{Data sets} \label{Sec:data_set}
In this study, we consider the following data sets:
\begin{itemize}
	
	\item CMB$_{\rm base}$: Planck 2018 low-l ({\tt Commander}) and high-l TT, TE, EE temperature power spectra ({\tt Plik}), supplemented by low-l EE polarization spectra ({\tt SimAll}) of photon anisotropies from CMB
	\item CMB$_{\rm full}$: same as CMB$_\text{base}$ with Planck 2018 conservative lensing spectrum added
	\item BAO: baryon acoustic oscillation measurements of large-scale structure from 6DF galaxy survey~\cite{Beutler_2011} at $z=0.106$, main galaxy sample (MGS) from SDSS DR7 survey~\cite{Ross:2014qpa} at $z=0.15$. We also consider the consensus data from BOSS DR12 SDSS survey~\cite{BOSS:2016wmc} at $z=0.38, 0.51, 0.61$, including the redshift-space distortion~(RSD) constraints on the structure growth $f\sigma_8$.
	\item WL: cosmic shear weak lensing data from a joint analyis of Kilo-Degree Survey (KV450) and Dark Energy Survey (DES-Y1)~\cite{Asgari:2019fkq}. For simplicity, we apply this data set in terms of a split-normal prior on $S_8 \equiv \sigma_8 (\Omega_m/0.3)^{0.5}$, such that
	\begin{align}
	S_8 = 0.755^{+0.019}_{-0.021}, \label{Eq:S8_prior}
	\end{align}
	which has been proven by~\cite{Hill:2020osr} to be equivalent to the computationally-expensive analysis of galaxy two-point correlations.
	\item SN: the Pantheon~\cite{Scolnic:2017caz} data set of 1048 supernovae Type Ia in the redshift range of $0.01 < z < 2.3$, which is given in terms of luminosity distances.
	\item R19: a Gaussian prior of the local Hubble parameter~\cite{Riess:2019cxk} from the recent SH0ES measurement using distance ladder method with Cepheid variables as calibrators
	\begin{align}
	H_0 = 74.03 \pm 1.42~\text{km/s/Mpc}. \label{Eq:H0_prior}
	\end{align}		
\end{itemize}

\subsection{Methodology} \label{Sec:methodology}

We apply Markov chain Monte Carlo (MCMC) analysis with {\tt CosmoMC}~\cite{Lewis:2002ah, Lewis:2013hha} for the following models:
\begin{itemize}
	\item $\Lambda$CDM: the standard cosmological model with six varying parameters 
	\begin{align}
	\{ \omega_b, \omega_c, \theta_\text{MC}, \tau_\text{reio}, \ln(10^{10}A_s), n_s\};
	\end{align}
	\item $\Lambda$CDM+$m_e$: an extension of $\Lambda$CDM with the electron mass $m_e$ varying constantly throughout the cosmic history as in~\cite{Hart:2017ndk}
	\begin{align}
	\{ \omega_b, \omega_c, \theta_\text{MC}, \tau_\text{reio}, \ln(10^{10}A_s), n_s, {m_e/m_{e,0}} \};
	\end{align}
	\item $\Lambda$CDM+$\omega_a$: an extension of $\Lambda$CDM with an ultra-light axion as in~\cite{Hlozek:2014lca}. We implicitly choose $m_a = 10^{-29}$ eV as the reference value, otherwise specifically stated.
	\begin{multline}
	\{ \omega_b, \omega_c, \theta_\text{MC}, \tau_\text{reio}, \ln(10^{10}A_s), n_s, \omega_a \} \\ \text{ with } m_a = 10^{-29}~\text{eV};
	\end{multline}
	
	\item Axi-Higgs: an extension of $\Lambda$CDM with an ultra-light axion driving the Higgs-VEV (or $m_e$) 
	\begin{multline}
	\{ \omega_b, \omega_c, \theta_\text{MC}, \tau_\text{reio}, \ln(10^{10}A_s), n_s, (v/v_0)_\text{ini}, \omega_a \} \\  \text{ with } m_a = 10^{-29}~\text{eV}. \label{Eq:params}
	\end{multline}
	Notice that axi-Higgs with the axion mass $m_a \lesssim 10^{-29}$ eV is phenomenologically equivalent to $\Lambda$CDM+$m_e$+$\omega_a$ model as mentioned above. The only distinction is that we set the electron mass nearly constant initially, i.e. $(v/v_0)_\text{ini} = m_e/m_{e,0}$, and let it precisely trace the axion background evolution afterward. App.~\ref{App:ah_lcdmmeax} examines the similarity as well as subtle difference between these two models.
\end{itemize}

Uniform priors are assumed on the varying cosmological parameters, particularly for the new parameters as follows
\begin{align}
0.9 < (v/v_0)_\text{ini} < 1.1, \quad 0 < \omega_a < 0.1.
\end{align}
The chains converge under Gelman-Rubin criterion where $R-1 < 0.03$ with the first 30\% steps are discarded as the burn-in phase. Marginalized posterior distributions and their plots are generated by {\tt GetDist}~\cite{Lewis:2019xzd}. The CMB lensing spectrum and the matter power spectrum are computed with non-linear corrections although the default {\tt halofit}~\cite{Takahashi:2012em} module used by {\tt CAMB} is tuned for the halo model in pure CDM cosmology~\footnote{Ref.~\cite{Hlozek:2016lzm} proved that the {\tt halofit} effect is small for axions with mass up to $10^{-26}$ eV. However, non-linear corrections of CDM-like axions require a dedicated code, e.g. {\tt WARMANDFUZZY}~\cite{Marsh:2016vgj}}. Tab.~\ref{Tab:ahcamb_test} in App.~\ref{App:supplementary} tests the reliability of {\tt aHCAMB} in reproducing the previous constraints of parameters in the $\Lambda$CDM+$m_e$ and $\Lambda$CDM+$\omega_a$ model.

\subsection{Results and Discussion} \label{Sec:models_compare}

We perform a joint analysis of CMB$_\text{full}$, BAO, WL, SN and R19 for the aforementioned models: $\Lambda$CDM, $\Lambda$CDM+$\omega_a$, $\Lambda$CDM+$m_e$ and axi-Higgs. The marginalized constraints of cosmological parameters are shown in Tab.~\ref{Tab:models_compare_full} and the posterior distributions can be found in Fig.~\ref{Fig:models_compare_full} in App.~\ref{App:supplementary}.

Firstly, we find that parameter constraints for $\Lambda$CDM obtained with this data combination are slightly different from the ones with CMB-only data. In particular, $H_0 = 68.57 \pm 0.37$ is found higher than $H_0 = 67.36 \pm 0.54$ by Planck~\cite{Aghanim:2018eyx}. The shift comes from two factors: (i) the $S_8$ prior of WL data that drags down $S_8$, as decreasing $S_8$ is associated with increasing $H_0$; (ii) the $H_0$ prior of R19 data that further raises $H_0$. Therefore, the inclusion of WL and R19 data naturally leads to a preference of higher $H_0$.

Secondly, we notice that the axion density of the $\Lambda$CDM+$\omega_a$ model is strictly bounded, $\omega_a \lesssim 0.001$, which implies no preference for substantial axion abundance from current data resolution. With a vanishingly small value of $\omega_a$, the $\Lambda$CDM+$\omega_a$ is indistinguishable from $\Lambda$CDM. Therefore, it is expected that other parameter constraints for $\Lambda$CDM+$\omega_a$, except $\omega_a$, are almost the same as their constraints for $\Lambda$CDM. Even the $\chi^2$ values of these two models are virtually identical, as seen in Tab.~\ref{Tab:models_compare_full}.

On the other hand, we find a significant improvement of $H_0$ in the $\Lambda$CDM+$m_e$ and axi-Higgs model with $H_0 = 70.75 \pm 0.90$ and $H_0 = 71.34 \pm 0.96$, respectively. Additionally, we see that the Higgs-VEV (electron mass) values are preferred above unity at roughly $1.4\%$ and $2.5\%$ with a noticeable increase of $\omega_b$ and $\omega_c$. These constraints convincingly illustrate the correlation between $H_0, (v/v_0)_{\rm ini}~({\rm or}~ m_e/m_{e,0})$ and other parameters in consequence of fixing the position of CMB acoustic peaks, as discussed in Sec.~\ref{Sec:cmb_pheno} Most notably, the bound of $\omega_a$ for axi-Higgs is more relaxed than $\Lambda$CDM+$\omega_a$ with a slight peak at $\omega_a \simeq 0.001$. The $H_0$ constraint for axi-Higgs also better alleviates the Hubble tension than $\Lambda$CDM+$m_e$, which seems counterintuitive since $\omega_a$ and $H_0$ are anti-correlated.

This result can be explained as follows. We know that $m_e$ and $H_0$ are positively degenerate with CMB data. The addition of R19 data breaks this degeneracy with its $H_0$ prior, so $\omega_b$ and $\omega_c$ also increase in order to preserve CMB acoustic structure. However, the simultaneous rise of $m_e, H_0, \omega_b, \omega_c$ may not preserve BAO angular scales at late times, which leads to a worse fit with BAO data. Therefore, the $\chi^2$ value of $\Lambda$CDM+$m_e$ is noticeably larger than $\Lambda$CDM and $\Lambda$CDM+$\omega_a$, at $\chi_{\rm BAO}^2 = 9.2$ compared to $\chi_{\rm BAO}^2 = 6.1$, respectively. As the Higgs-VEV is uplifted, the axi-Higgs model faces the same issue with BAO data, but to a lesser extent thanks to the presence of the $10^{-29}~{\rm eV}$ axion. The DE-like nature of this axion offers one more degree of freedom to modify the low-z expansion rate without affecting the BAO sound horizon~\footnote{Non-flat universe with varying $\Omega_k$ give rises to similar effects, as proposed by the $\Lambda$CDM+$m_e$+$\Omega_k$ model~\cite{Sekiguchi:2020teg}}. Thus, parameter constraints for axi-Higgs yield a higher value of $H_0$ and a higher amount of $\omega_a$ compared to $\Lambda$CDM+$m_e$ while having an enhanced BAO fit with $\chi_{\rm BAO}^2 = 7.7$. The preferred $H_0$ and $\omega_a$ values also improve R19 fit and worsen CMB lensing fit. Eventually, the Hubble tension is alleviated in the axi-Higgs model with an overall value of $\chi^2 = 3837.4$, which is significantly lower than $\Lambda$CDM at $\chi^2 = 3843.8$, and comparable to $\Lambda$CDM+$m_e$ at $\chi^2 = 3836.6$. The small difference between the best-fit $\chi^2$ of axi-Higgs and $\Lambda$CDM+$m_e$ is due to numerical artifacts as commented in the description of Tab.~\ref{Tab:models_compare_full}.

For further comparison, we provide more parameter constraints of these four models in App.~\ref{App:supplementary} with different data combinations, particularly CMB$_{\rm base}$+BAO in Tab.~\ref{Tab:models_compare_nowl} and CMB$_{\rm base}$+BAO in Tab.~\ref{Tab:models_compare_base}. Most notably, even without the inclusion of SN and SH0ES data $H_0$ is still improved up to $2.4\sigma$ at $H_0 = 69.4 \pm 1.3$ with CMB$_{\rm base}$+BAO+WL data in the axi-Higgs model compared to the $\Lambda$CDM model. This $H_0$ constraint is also superior to the one found in the $\Lambda$CDM$+m_e$ model at $H_0 = 68.6 \pm 1.1$, which further proves the important role of the ultralight axion as discussed above.

\begin{table*}[htp]
		\begin{tabular}{c|c|c|c|c}
			
			Data & \multicolumn{4}{c}{CMB$_\text{full}$+BAO+WL+SN+R19}  \\
			\hline
			Model & $\Lambda$CDM & $\Lambda$CDM+$\omega_a$ & $\Lambda$CDM+$m_e$ & axi-Higgs \\
			\hline\hline
			
			$\omega_b$ & $0.02259\,(0.02261) \pm 0.00013$ & $0.02260\,(0.02261) \pm 0.00013$ & $0.02276\,(0.02275) \pm 0.00015$ & $0.02293\,(0.02294) \pm 0.00017$ \\
			$\omega_c$ & $0.11737\,(0.11730) \pm 0.00082$ & $0.11723\,(0.11735) \pm 0.00080$ & $0.1211\,(0.1211) \pm 0.0016$ & $0.1235\,(0.1235)^{+0.0020}_{-0.0023}$ \\
			$100\theta_\text{MC}$ & $1.04120\,(1.04120) \pm 0.00029$ & $1.04123\,(1.04122) \pm 0.00028$ & $1.0511\,(1.0511) \pm 0.0037$ & $1.0586\,(1.0585)^{+0.0052}_{-0.0064}$ \\
			$\tau_\text{reio}$ & $0.0569\,(0.0570)^{+0.0066}_{-0.0073}$ & $0.0579\,(0.0579)^{+0.0065}_{-0.0076}$ & $0.0509\,(0.0510) \pm 0.0073$ & $0.0552\,(0.0552) \pm 0.0077$ \\
			$\ln (10^{10} A_s)$ & $3.044\,(3.044) \pm 0.014$ & $3.046\,(3.046) \pm 0.014$ & $3.036\,(3.036) \pm 0.014$ & $3.049\,(3.049) \pm 0.016$ \\
			$n_s$ & $0.9708\,(0.9720) \pm 0.0036$ & $0.9712\,(0.9719) \pm 0.0036$ & $0.9660\,(0.9666) \pm 0.0041$ & $0.9649\,(0.9649) \pm 0.0040$ \\
			\hline
			$(v/v_0)_\text{ini}$ & $1$ & $1$ & $1.0143\,(1.0144) \pm 0.0053$ & $1.0254\,(1.0251)^{+0.0076}_{-0.0093}$ \\
			$100\omega_a$ & $0$ & $< 0.0623\,(\sim 0)$ & $0$ & $ 0.140\,(0.142)^{+0.052}_{-0.12}$ \\
			\hline
			$H_0$ & $68.57\,(68.61) \pm 0.37$ & $68.42\,(68.59) \pm 0.40$ & $70.75\,(70.73) \pm 0.90$ & $71.34\,(71.24) \pm 0.96$ \\
			$S_8$ & $0.8019\,(0.8013) \pm 0.0087$ & $0.7998\,(0.8026) \pm 0.0087$ & $0.8049\,(0.8050) \pm 0.0087$ & $0.8017\,(0.8017) \pm 0.0093$ \\
			$\sigma_8$ & $0.8031\,(0.8032) \pm 0.0055$ & $0.7990\,(0.8041)^{+0.0066}_{-0.0058}$ & $0.8205\,(0.8205) \pm 0.0083$ & $0.813\,(0.8119) \pm 0.010$ \\
			$\phi_{\rm ini}$ & $0$ & $1.10\,(\sim 0)^{+0.36}_{-0.63}$ & $0$ & $2.73\,(2.92)^{+1.1}_{-0.86}$ \\
			\hline\hline
			$\chi^2_\text{tot}$ & $3843.8$ & $3844.0$ & $3836.6$ & $3837.4$ \\
			$\Delta \chi^2_\text{tot}$ & $0$ & $0.2$ & $-7.2$ & $-6.4$ \\
			\hline
			$\chi^2_\text{TTTEEE high-l}$ & $2350.4$ & $2350.2$ & $2346.7$ & $2350.2$ \\
			$\chi^2_\text{TT low-l}$ & $22.2$ & $22.2$ & $23.2$ & $23.7$ \\
			$\chi^2_\text{EE low-l}$ & $396.3$ & $396.6$ & $395.7$ & $396.1$ \\
			$\chi^2_\text{lensing}$ & $9.8$ & $9.6$ & $9.42$ & $10.1$ \\
			\hline
			$\chi^2_\text{6dF}$ & $\sim 0$ & $\sim 0$ & $0.2$ & $0.1$ \\
			$\chi^2_\text{MGS}$ & $2.2$ & $2.2$ & $3.2$ & $2.8$ \\
			$\chi^2_\text{BAO DR12}$ & $6.1$ & $6.1$ & $9.2$ & $7.7$ \\
			\hline
			$\chi^2_\text{WL}$ & $5.9$ & $6.3$ & $6.9$ & $6.1$ \\
			\hline
			$\chi^2_\text{SN}$ & $1034.7$ & $1034.7$ & $1035.0$ & $1034.8$ \\
			\hline
			$\chi^2_\text{R19}$ & $14.6$ & $14.7$ & $5.4$ & $3.9$
		\end{tabular}
	\caption{
	Marginalized distributions of parameters in four models fitted with CMB$_\text{full}$+BAO+WL+SN+R19. The uncertainties and the upper bounds of parameters are shown at 68\% and 95\% confidence levels, respectively. Note that $(v/v_0)_{\rm ini} = m_e/m_{e,0}$ in the $\Lambda$CDM+$m_e$ model. The best-fit values are quoted in the parentheses next to the mean values, which is found by {\tt BOBYQA} minimization routine~\cite{Powell2009TheBA} embedded in {\tt CosmoMC}. The $\chi^2$ statistics are calculated for the best-fit parameters of each model, where $\Delta \chi_{\rm tot}^2$ indicates $\chi_{\rm tot}^2$ difference between the corresponding model and $\Lambda$CDM. Since the axion dynamics is activated only when $\omega_a > 10^{-5}$ in \texttt{aHCAMB} to avoid numerical instability, {\tt BOBYQA} struggles to search for the global minimum of cosmological models with an extended parameter space by small but finite values of $\omega_a$. Therefore, the reported $\chi^2$ in $\Lambda$CDM+$m_e$ is slightly better than axi-Higgs, and similarly the $\chi^2$ of $\Lambda$CDM is slightly better than $\Lambda$CDM+$\omega_a$ even though the former models are more general than the latter models.}
	\label{Tab:models_compare_full}
\end{table*}

\begin{figure}[!ht]
	\centering
	\includegraphics[scale=0.43]{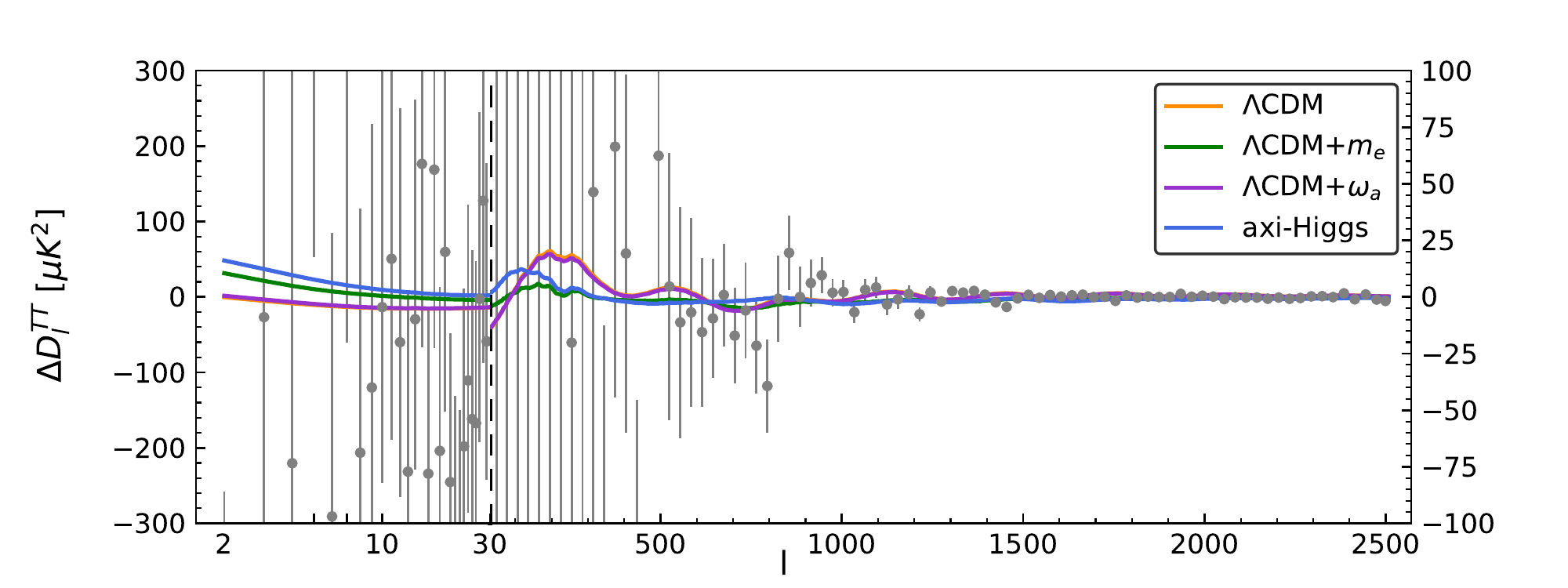}
	\includegraphics[scale=0.43]{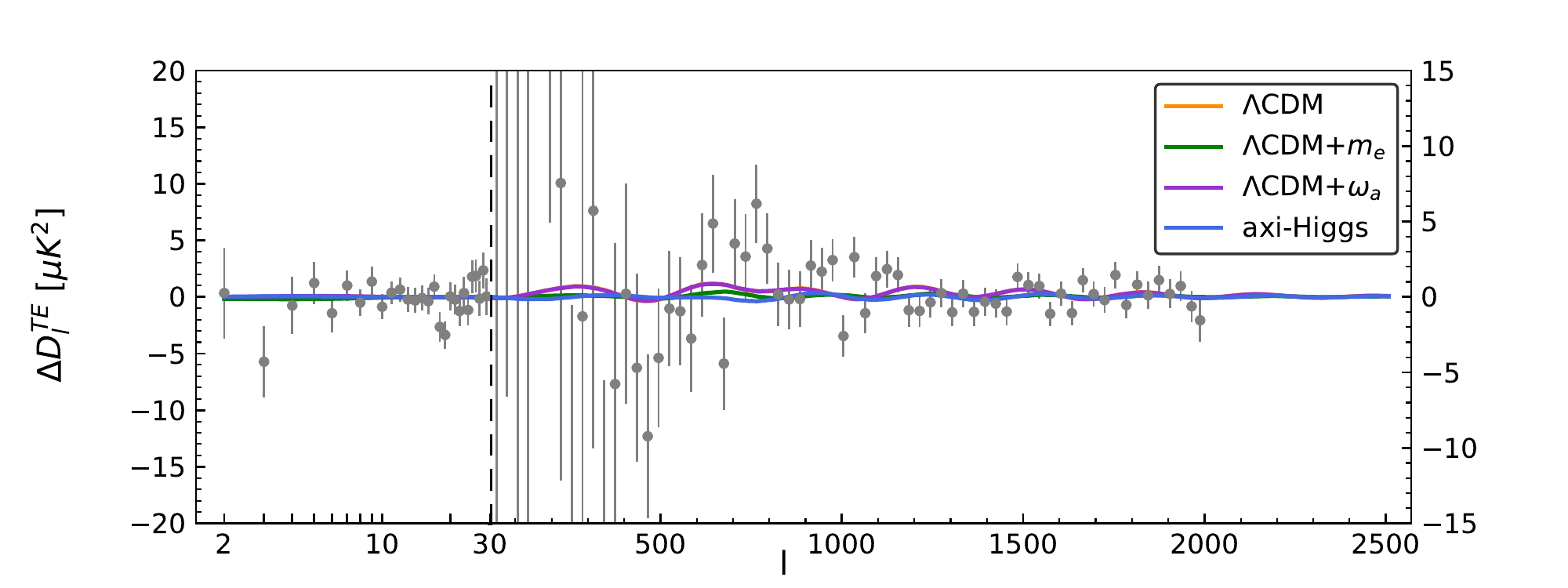}
	\includegraphics[scale=0.43]{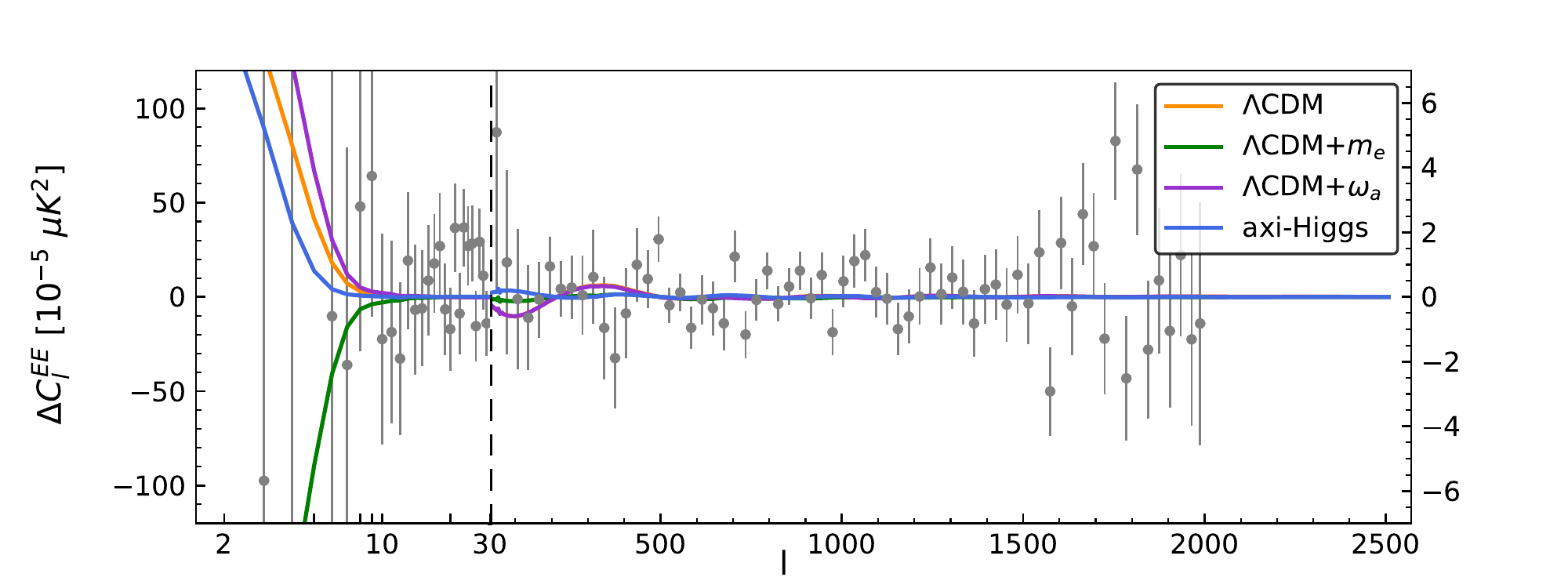}
	\includegraphics[scale=0.43]{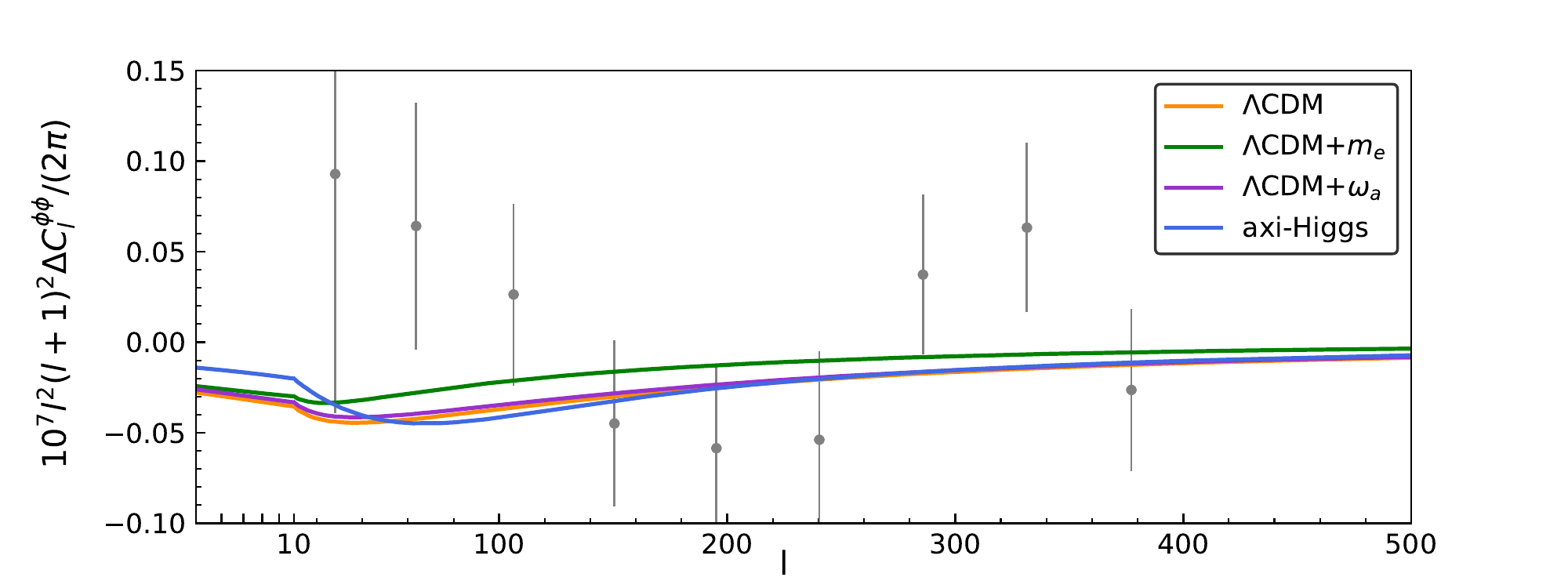}
	\caption{Residuals of TT, TE, EE, $\phi\phi$ spectra with respect to CMB data in $\Lambda$CDM, $\Lambda$CDM+$m_e$, $\Lambda$CDM+$\omega_a$, axi-Higgs. We define $\Delta C_l^{X} \equiv C^{X}_{l, Y} - C^{X}_{l, {\rm fid}}$ where $X$'s are spectral indices and $Y$'s are model indices. $C_{l, Y}$ spectra are computed with the best-fit parameters from Tab.~\ref{Tab:models_compare_full} while the reference spectrum $C_{l, {\rm fid}}$ is obtained with the best fit of $\Lambda$CDM with CMB$_{\rm full}$, as provided by Planck 2018~\cite{Aghanim:2018eyx}. The dashed vertical lines at $l=30$ separate the low-l and high-l region. The low-l TE data points are shown for illustration but not included in our data fitting.}
	\label{fig:aH_residuals}
\end{figure}


\section{Conclusion and Outlook} \label{Sec:conclusion}

In this work, we have comprehensively studied the cosmological implications of the axi-Higgs model, where the Higgs-VEV is driven by an ultralight axion with mass $m_a \sim 10^{-29}$~eV. Under presently available data from CMB, LSS and local astrophysical measurements, we place constraints on the cosmological parameters of the axi-Higgs model and make systematic comparison with other models, including $\Lambda$CDM, $\Lambda$CDM+$m_e$, $\Lambda$CDM+$\omega_a$. The results suggest that the axi-Higgs model potentially resolves the Hubble tension to restore the cosmic concordance. We note that similar conclusion has been reached in~\cite{Fung:2021fcj}, independently with the leading-order perturbative approach (LPA).

There remain many topics that we have not elaborated on from a broader perspective of the axi-Higgs model. Firstly, on the subject of BBN, Ref.~\cite{Fung:2021wbz} has predicted that the Higgs-VEV must stay roughly 1\% above its present-day value to alleviate the $^7$Li puzzle~\cite{Fields:2011zzb}, which turns out to be consistent with the $(v/v_0)_{\rm ini}$ constraint obtained in Tab.~\ref{Tab:models_compare_base}. Secondly, the interaction term of axion and photon $gF\tilde{F}$, which has been ignored so far, is responsible for the rotation of the photon polarization plane~\cite{Harari:1992ea}. Consequently, it may induce a non-vanishing isotropic cosmic birefringence~(ICB) signal on CMB data with the constraint of the ICB angle $\beta = 0.35 \pm 1.4~{\rm deg}$~\cite{Minami:2020odp}, which is converted to $f_a \sim 10^{17}-10^{18}$~GeV for $\omega_a \sim 0.001$, see~\cite{Fujita:2020aqt, Fung:2021wbz} for more details. Finally, the Higgs field can couple to the second axion field with a higher mass~\cite{Fung:2021wbz}, i.e. $\delta v = (C_1\phi_1^2 + C_2\phi_2^2)/2 M^2_\text{pl}$ from theoretical perspectives. In that case, the parameter $(v/v_0)_{\rm ini}$ used so far should be interpreted as the Higgs-VEV deviation at recombination $(v/v_0)_{\rm rec}$ since the Higgs-VEV deviation at BBN $(v/v_0)_{\rm BBN}$ can be set to a different value. The two-axion model is necessary to archive the 2\% Higgs-VEV uplift as reported in Tab.~\ref{Tab:models_compare_full} at recombination and the 1\% deviation at BBN as required by the produced abundance of light elements~\cite{Fung:2021wbz}. This extended framework could simultaneously resolve the $^7$Li puzzle, the Hubble and $S_8$ tension, and explain small-scale crisis if the heavy axion is FDM with $m_a \sim 10^{-22}$~eV~\cite{Hu:2000ke}. An extensive treatment for these interesting questions will be the main theme of our future studies.

\begin{acknowledgments}
I would like to express my appreciation to Prof.~Henry Tye and Prof.~Tao Liu for the initial ideas leading to this paper. I thank my colleagues Leo Fung, Lingfeng Li and Yu-Cheng Qiu for useful discussions on the related topics. I am grateful for the support and encouragement from my dear Shuting during the time of writing this paper. This work is supported by the Area of Excellence under Grant No.~AoE/P-404/18-3(6) issued by the Research Grants Council of Hong Kong SAR.
\end{acknowledgments}

\appendix

\section{Axi-Higgs with different axion mass} \label{App:aH_compare}

\begin{figure*}[b]
	\centering
	\includegraphics[scale=0.5]{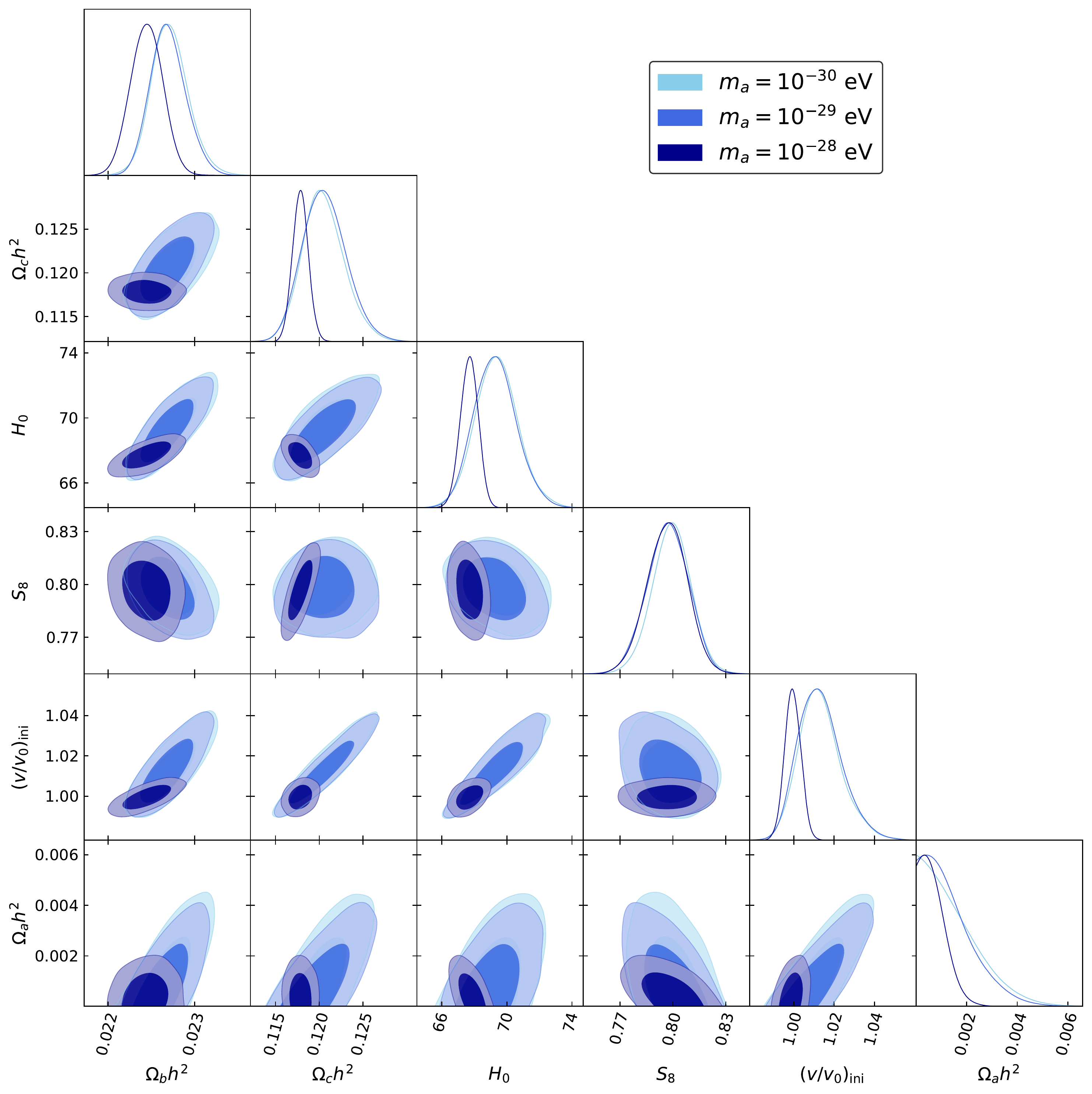}
	\caption{Posterior distributions of cosmological parameters for axi-Higgs fitted with CMB$_{\rm base}$+BAO+WL with different axion masses. The 1D and 2D marginalized contours of the $10^{-30}$-eV axion are behind the contours of the $10^{-29}$-eV axion.}
	\label{Fig:aH_mass_compare}
\end{figure*}

As the Higgs-VEV deviation is governed by the axion evolution, it is ideal for the axion mass to lie within $10^{-30} \leq m_a \leq 10^{-29}$~\cite{Fung:2021wbz} to maintain a significant uplift of the Higgs-VEV before recombination. An axion heavier than $10^{-29}~{\rm eV}$ would yield negligible deviation of $m_e$ by recombination since its squared amplitude starts damping at earlier times. An axion lighter than $10^{-30}~{\rm eV}$ is excluded by quasar observations~\cite{Levshakov:2020ule} and atomic clock experiments~\cite{Lange:2020cul} since its amplitude is not sufficiently attenuated by the present day. Therefore, we explore parameter constraints of the axi-Higgs model with different axion masses: $10^{-28}~{\rm eV}, 10^{-29}~{\rm eV}, 10^{-30}~{\rm eV}$ in this section. The marginalized constraints and posterior distributions are shown  in Tab.~\ref{Tab:aH_mass_compare} and Fig.~\ref{Fig:aH_mass_compare}, respectively.
	
	As expected, we find that the parameter constraints from the $10^{-28}$-eV axion in the axi-Higgs model are mostly the same as $\Lambda$CDM+$\omega_a$ in Sec.~\ref{Sec:models_compare}. Since this axion transits at $z_\text{osc} \simeq 1090$, barely before recombination at $z_\text{osc} \simeq 1100$, its amplitude is damped by roughly one order of magnitude from an initial deviation at recombination. For instances, $(v/v_0)_{\rm ini} = 1\%$ would become $(v/v_0)_{\rm rec} = 0.1\%$, which negligibly impacts recombination physics. Thus, the axi-Higgs model essentially converges to $\Lambda$CDM+$\omega_a$ model as the axion mass becomes heavier. More importantly, this result shows that the initial uplift of the Higgs-VEV should remain (almost) constant in a narrow window of redshift around recombination to distinguish the axi-Higgs model from the standard regime of axion cosmology.
	
	On the other hand, we find that the parameter constraints from the $10^{-30}$-eV axion resemble the $10^{-29}$-eV counterparts. We also notice in Fig.~\ref{Fig:aH_mass_compare} that the peak of $\omega_a$ distribution is shifted towards zero and its upper bound is slighted relaxed as $m_a$ decreases. This trend is anticipated to continue for even lighter axions with $m_a < 10^{-30}~{\rm eV}$ as they tend to be indistinguishable with DE. In that case, the axi-Higgs and $\Lambda$CDM+$m_e$ model inevitably converge.
	
	Thus, the results show that the distinct features of the axi-Higgs model only manifest themselves when the axion mass falls within $10^{-30}~{\rm eV} < m_a < 10^{-29}~{\rm eV}$.

\begin{table*}[htp]
	\begin{tabular}{c|c|c|c}
		Model & \multicolumn{3}{c}{axi-Higgs}  \\
		\hline
		Data & \multicolumn{3}{c}{CMB$_\text{base}$+BAO+WL}  \\
		\hline
		$m_a$ & $10^{-28}$ eV & $10^{-29}$ eV & $10^{-30}$ eV \\
		\hline\hline
		
		$\omega_b$ & $0.02245 \pm 0.00018$ & $0.02270^{+0.00019}_{-0.00022}$ &  $0.02271^{+0.00019}_{-0.00023}$ \\
		$\omega_c$ & $0.11785 \pm 0.00090$  &  $0.1207^{+0.0022}_{-0.0027}$ & $0.1205^{+0.0021}_{-0.0025}$ \\
		$100\theta_\text{MC}$ & $1.04103 \pm 0.00092$ & $1.0505^{+0.0061}_{-0.0083}$ & $1.0503^{+0.0060}_{-0.0081}$ \\
		$\tau_\text{reio}$ & $0.0527 \pm 0.0077$ & $0.0527 \pm 0.0079$ & $0.0521 \pm 0.0077$ \\
		$\ln (10^{10} A_s)$ & $3.034 \pm 0.016$ & $3.038 \pm 0.016$ & $3.037 \pm 0.016$ \\
		$n_s$ & $0.9673 \pm 0.0056$ & $0.9672 \pm 0.0042$ & $0.9670 \pm 0.0043$ \\
		\hline
		$(v/v_0)_\text{ini}$ & $0.9995 \pm 00040$ & $1.0135^{+0.0088}_{-0.012}$ & $1.0131^{+0.0086}_{-0.012}$ \\
		$\omega_a$ & $< 0.00164$ & $< 0.00323$ & $ < 0.00362$ \\
		\hline
		$H_0$ & $67.69 \pm 0.54$ & $69.4 \pm 1.3$ & $69.4 \pm 1.3$ \\
		$S_8$ & $0.797 \pm 0.011$ & $0.797 \pm 0.012$ & $0.799 \pm 0.011$ \\
		$\sigma_8$ & $0.785^{+0.013}_{-0.0099}$ & $0.794 \pm 0.013$ & $0.796 \pm 0.012$ \\
		$\phi_{\rm ini}$ & $1.69 \pm 0.65$ & $2.6 \pm 1.1$ & $2.9 \pm 1.2$
	\end{tabular}
	\caption{Marginalized distributions of parameters in axi-Higgs fitted with CMB$_{\rm base}$+BAO+WL for different axion masses.}
	\label{Tab:aH_mass_compare}
\end{table*}

\section{The (non-)equivalence \\ of axi-Higgs and $\Lambda$CDM+$m_e$+$\omega_a$} \label{App:ah_lcdmmeax}

The axi-Higgs model is equivalent but not identical to the $\Lambda$CDM+$m_e$+$\omega_a$ model. The latter is an extension of $\Lambda$CDM which allows a non-standard electron mass with the addition of an axion. Thus, the $m_e$ deviation in this model is always constant, independent of axion dynamics. On the contrary, the electron mass in the axi-Higgs model is modulated by the Higgs-VEV variation. Thus, the initial $m_e$ deviation will decay as soon as the axion density starts diluting at late times, see Fig.~\ref{Fig:aH_efa}. Depending on the cosmic epoch being considered, the main distinctive physics of axi-Higgs with $\Lambda$CDM+$m_e$+$\omega_a$ include: (i)~BBN when the Higgs-VEV uplift alters the abundance of several primordial elements such as D, $^4$He, $^7$Li; (ii)~reionzation when the electron mass deviation is suppressed compared to its initial value. It is worth noticing that cosmological models which change the electron mass without dismissing its deviation would be in conflict with astrophysical and laboratory observations at late times~\cite{Fung:2021wbz}. The axi-Higgs model converges to: $\Lambda$CDM when $\omega \rightarrow 0$; $\Lambda$CDM+$\omega_a$ when $(v/v_0)_{\rm ini} \rightarrow 1$; $\Lambda$CDM+$m_e$ when $m_a \rightarrow 0$ whereas the $\Lambda$CDM+$m_e$+$\omega_a$ model converges to: $\Lambda$CDM when $\omega_a \rightarrow 0$ and $(m_e/m_{e,0}) \rightarrow 1$; $\Lambda$CDM+$\omega_a$ when $(m_e/m_{e,0}) \rightarrow 1$; $\Lambda$CDM+$m_e$ when $\omega_a \rightarrow 0$.

\begin{figure*}[b]
	\centering
	\includegraphics[scale=0.5]{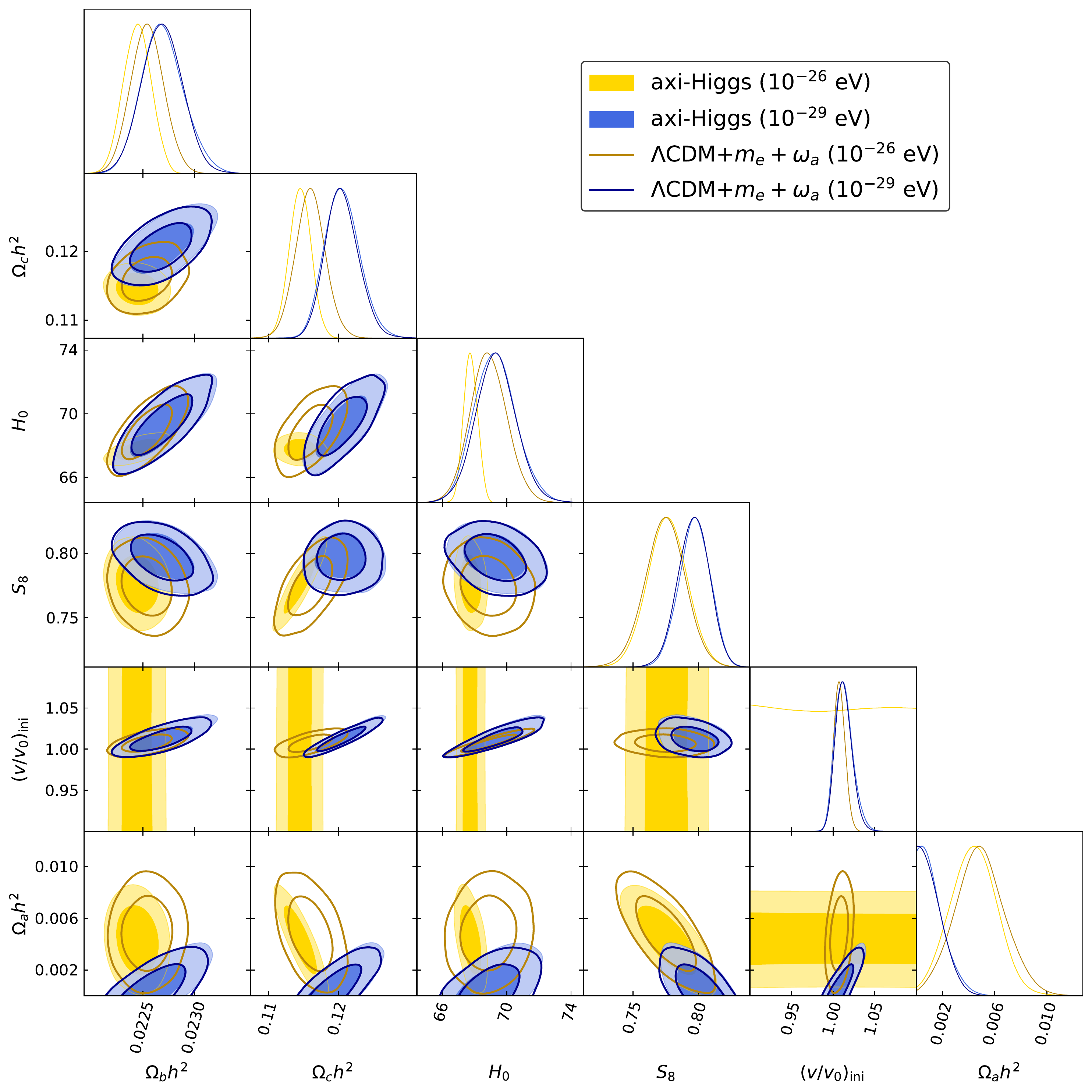}
	\caption{Posterior distributions of cosmological parameters for axi-Higgs and $\Lambda$CDM+$m_e$+$\omega_a$ fitted with CMB$_{\rm base}$+BAO+WL.}
	\label{Fig:aH_lcdmmeax_compare}
\end{figure*}

Despite some subtle difference, the $\Lambda$CDM+$m_e$+$\omega_a$ model can be perfectly treated as the phenomenological model of axi-Higgs for $m_a \lesssim 10^{-29}$ eV. We prove it with an analysis for these two models with parameter constraints shown in Tab.~\ref{Tab:aH_lcdmmeax_compare}. We clearly find that their constraints with $m_a = 10^{-29}~{\rm eV}$ are exactly identical. The corresponding posterior distributions, shown in Fig.~\ref{Fig:aH_lcdmmeax_compare}, also perfectly overlap.

However, a heavier axion will differentiate axi-Higgs and $\Lambda$CDM+$m_e$+$\omega_a$. The parameter constraints for these two models with $m_a = 10^{-26}$ eV look drastically different, especially with $(v/v_0)_{\rm ini}$ unbounded in the case of axi-Higgs. The reason is because the axion oscillation was triggered at very high redshifts, i.e. $z_{\rm osc} \sim \mathcal{O}(10^5)$, which makes any reasonable value of the initial Higgs-VEV deviation disappearing long before the epoch of recombination. We also notice that the axion density is preferred at $\omega_a \simeq 0.005$ for this heavy axion. Thus, $S_8$/$\sigma_8$ is substantially suppressed, which can be an important hint for resolving the $S_8$ tension. In the literature, the inclusion of the $10^{-26}~{\rm eV}$ axion reconciles the exceed DM density preferred by EDE-type models~\cite{Allali:2021azp, Ye:2021iwa}.

\begin{table*}[htp]
		\begin{tabular}{c|c|c|c|c}
			
			Data & \multicolumn{4}{c}{CMB$_\text{base}$+BAO+WL}  \\
			\hline
			$m_a$ & \multicolumn{2}{c|}{$10^{-29}$ eV} & \multicolumn{2}{c}{$10^{-26}$ eV} \\
			\hline
			Model & axi-Higgs & $\Lambda$CDM+$m_e$+$\omega_a$ & axi-Higgs & $\Lambda$CDM+$m_e$+$\omega_a$ \\
			
			\hline\hline
			
			$\omega_b$ & $0.02270^{+0.00019}_{-0.00022}$ & $0.02269 \pm 0.00020$ & $0.02244 \pm 0.00013$ & $0.02254 \pm 0.00016$ \\
			$\omega_c$ & $0.1207^{+0.0022}_{-0.0027}$ &  $0.1206^{+0.0021}_{-0.0025}$ & $0.1145 \pm 0.0016$ & $0.1160 \pm 0.0021$ \\
			$100\theta_\text{MC}$ & $1.0505^{+0.0061}_{-0.0083}$ & $1.0501^{+0.0061}_{-0.0075}$ & $1.04114 \pm 0.00029$ & $1.0463 \pm 0.0049$ \\
			$\tau_\text{reio}$ & $0.0527 \pm 0.0079$ & $0.0522 \pm 0.0079$ & $0.0545 \pm 0.0075$ & $0.0533 \pm 0.0078$ \\
			$\ln (10^{10} A_s)$ & $3.038 \pm 0.016$ & $3.037 \pm 0.016$ & $3.040 \pm 0.015$ & $3.040 \pm 0.016$ \\
			$n_s$ & $0.9672 \pm 0.0042$ & $0.9669 \pm 0.0043$ & $0.9642 \pm 0.0040$ & $0.9617 \pm 0.0048$ \\
			\hline
			$(v/v_0)_\text{ini}$ & $1.0135^{+0.0088}_{-0.012}$ & $1.0128^{+0.0086}_{-0.011}$ & unbounded & $1.0074 \pm 0.0070$ \\
			$\omega_a$ & $<0.00323$ & $<0.00304$ & $0.0044 \pm 0.0017$ & $0.0049 \pm 0.0019$ \\
			\hline
			$H_0$ & $69.4 \pm 1.3$ & $69.3 \pm 1.3$ & $67.75 \pm 0.43$ & $68.9 \pm 1.1$ \\
			$S_8$ & $0.797 \pm 0.012$ & $0.796 \pm 0.012$ & $0.776 \pm 0.015$ & $0.775 \pm 0.015$ \\
			$\sigma_8$ & $0.794 \pm 0.013$ & $0.793 \pm 0.013$ & $0.764 \pm 0.015$ & $0.770 \pm 0.016$ \\
			$\phi_{\rm ini}$ & $2.6 \pm 1.1$ & $2.6 \pm 1.0$ & $1.99^{+0.49}_{-0.33}$ & $2.10^{+0.48}_{-0.35}$
		\end{tabular}
	\caption{Marginalized constraints of parameters in axi-Higgs and $\Lambda$CDM+$m_e$+$\omega_a$ fitted with CMB$_\text{base}$+BAO+WL for $m_a = 10^{-29}$ eV and $m_a = 10^{-26}$ eV. The unbounded constraint of $(v/v_0)_{\rm ini}$ indicates a uniform posterior spanning its entire prior range, see Fig.~\ref{Fig:aH_lcdmmeax_compare}.}
	\label{Tab:aH_lcdmmeax_compare}
\end{table*}

\section{The validity of \\ effective fluid approximation} \label{App:EFA}

Integrating the exact axion equations is computationally expensive due to the mismatch of the intrinsic Compton period and the Hubble time scale. In practice, we can average over several periods of axion oscillations to derive the effective fluid approximation~(EFA) solution after the background field becomes dynamical as described in Sec.~\ref{Sec:ah_physics}. The question we explore in this section is when that transition is approximately triggered in EFA.

\begin{figure*}[!ht]
	\centering
	\includegraphics[scale=0.45]{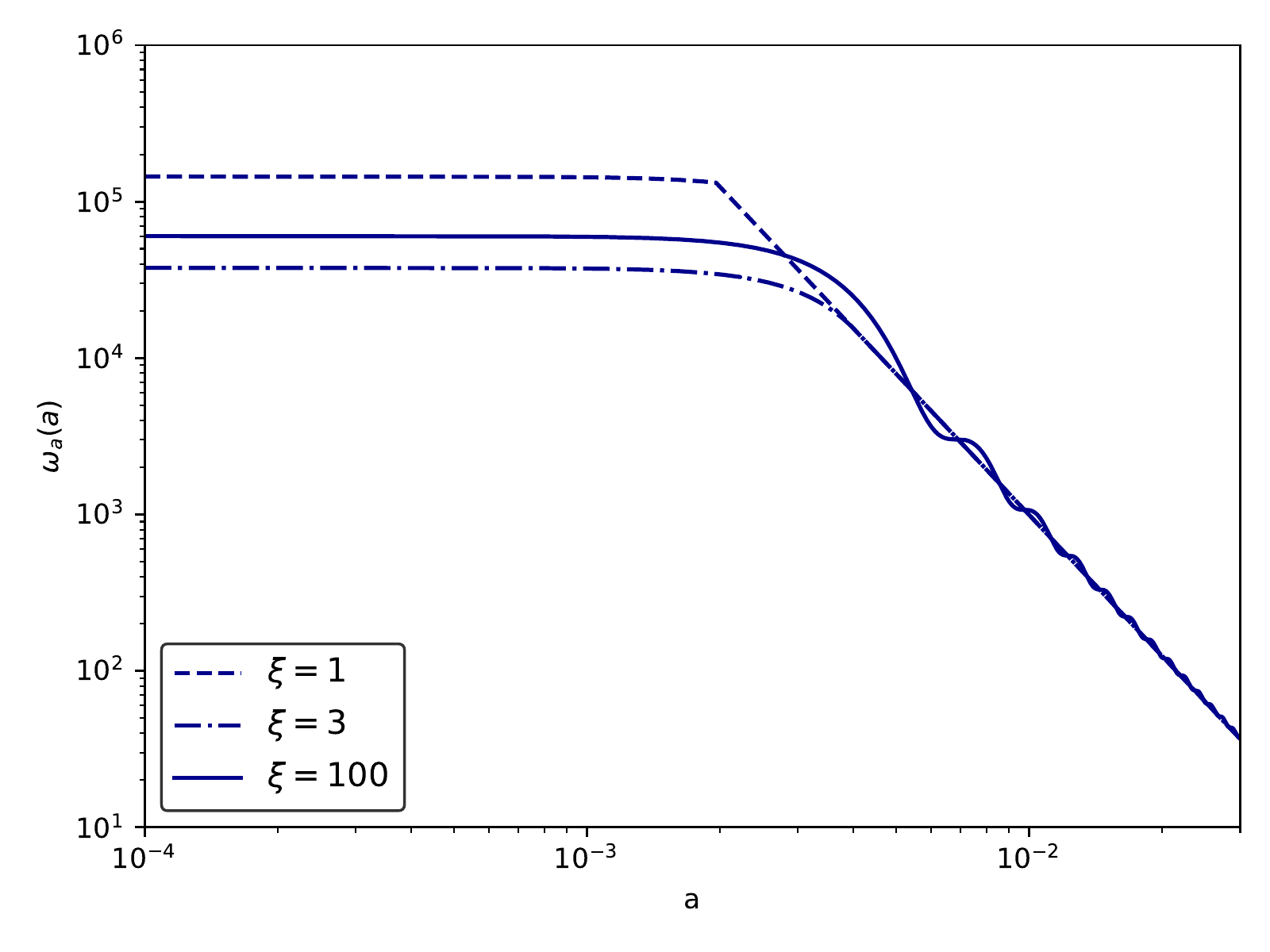}\includegraphics[scale=0.45]{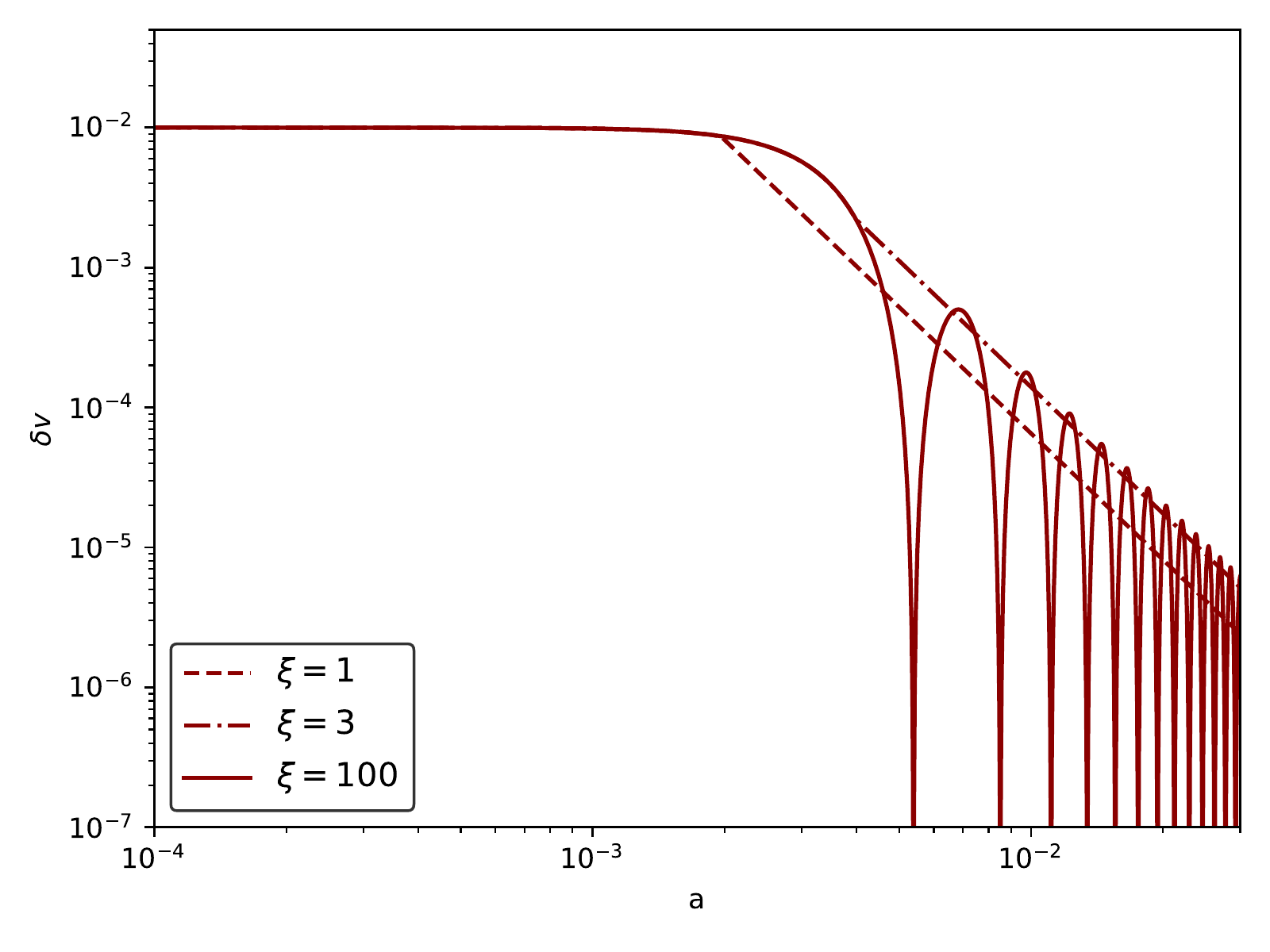} \\
	\includegraphics[scale=0.46]{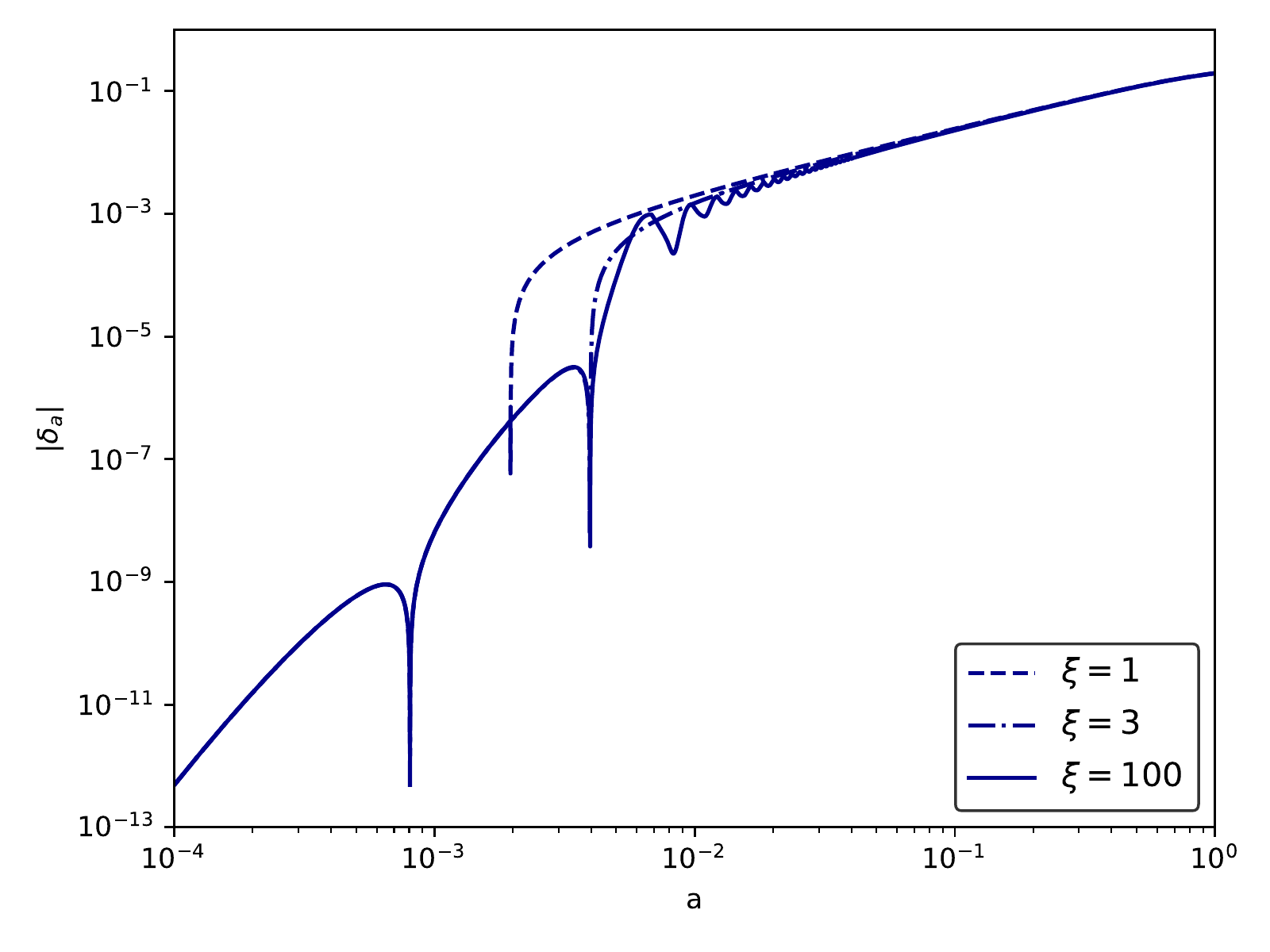}\includegraphics[scale=0.46]{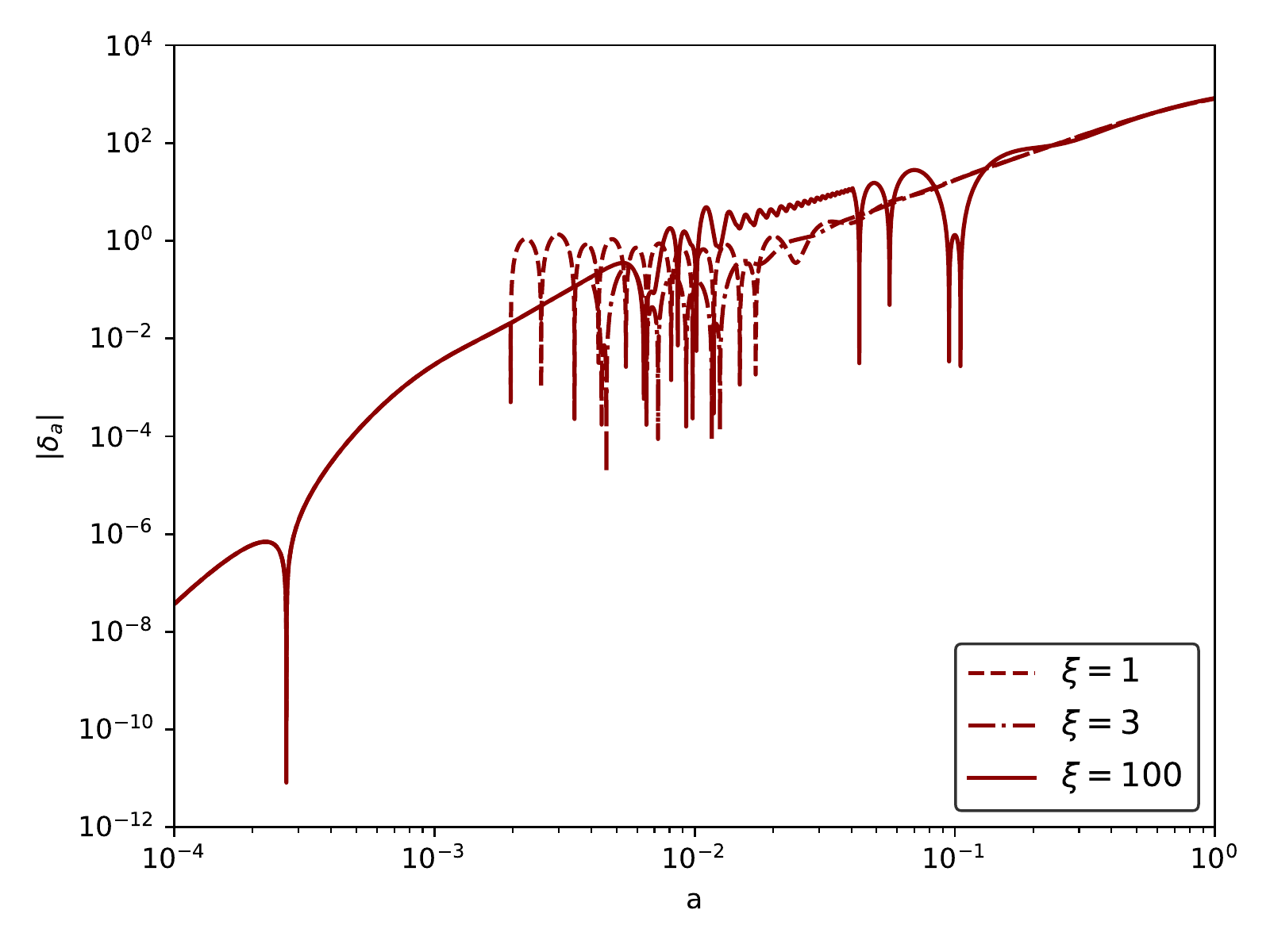}
	\caption{Axion dynamics with different transition coefficients in the axi-Higgs model. \textit{(Upper)} The axion background density on the left and the Higgs-VEV deviation on the right. \textit{(Lower)} The axion perturbations of $k = 10^{-4}~\text{Mpc}^{-1}$ on the left and $k = 0.03~\text{Mpc}^{-1}$ on the right. The plots are generated with cosmological parameters fixed to the fiducial values.}
	\label{Fig:aH_efa}
\end{figure*}

The axion equation of motion~\eqref{Eq:ax_eom} tells us that the harmonic term, i.e. $m_a^2a^2\phi$, is dominated by the Hubble friction until $H \sim m_a$, so this moment is a crude estimation of the transition redshift. We then parametrize this condition with the coefficient $\xi$ in Eq.~\eqref{Eq:ax_osc}, which is typically chosen to be $\xi = 3$ in the literature~\cite{Hlozek:2014lca, Marsh:2015xka}. In principle, the later the transition happens, i.e. the larger $\xi$ is, the more accurate dynamics will be captured. Fig.~\ref{Fig:aH_efa} demonstrates how physical quantities of the axi-Higgs model evolve given the common choice $\xi = 3$, the crude dynamics $\xi = 1$ or the nearly-exact dynamics $\xi = 100$. Based on the fact that the approximate curves closely trace the exact evolution in each plot, we are reassured that the $\xi = 3$ choice is sufficient up to necessary precision of the current data. However, EFA implementation with the switch at $\xi = 3$ may introduce biases on CMB spectra of more than $4\sigma$ for higher axion masses~\cite{Cookmeyer:2019rna}. \\

\section{Supplementary materials} \label{App:supplementary}

This appendix shows additional information quoted in the main text for completeness and self-consistency.

\begin{table*}[htp]
		\begin{tabular}{c|c|c|c|c}
			Data & \multicolumn{4}{c}{CMB$_\text{base}$+BAO}  \\
			\hline
			Model & $\Lambda$CDM & $\Lambda$CDM+$\omega_a$ & $\Lambda$CDM+$m_e$ & axi-Higgs  \\
			\hline\hline
			
			$\omega_b$ & $0.02241 \pm 0.00013$ & $0.02242 \pm 0.00014$ &  $0.02246 \pm 0.00017$ & $0.02254 \pm 0.00018$\\
			$\omega_c$ & $0.11940 \pm 0.00097$  & $0.1192 \pm 0.0010$ & $0.1205^{+0.0018}_{-0.0020}$ & $0.1215 \pm 0.0022$ \\
			$100\theta_\text{MC}$ & $1.04097 \pm 0.00028$ & $1.04101 \pm 0.00030$ & $1.0438 \pm 0.0046$ & $1.0476^{+0.0053}_{-0.0060}$ \\
			$\tau_\text{reio}$ & $0.0555^{+0.0070}_{-0.0080}$ & $0.0560 \pm 0.0082$ & $0.0543 \pm 0.0079$ & $0.0554 \pm 0.0078$ \\
			$\ln (10^{10} A_s)$ & $3.045 \pm 0.016$ & $3.046 \pm 0.017$ & $3.044 \pm 0.016$ & $3.038 \pm 0.016$ \\
			$n_s$ & $0.9661 \pm 0.0036$ & $0.9665 \pm 0.0038$ & $0.9648 \pm 0.0042$ & $0.9643 \pm 0.0042$ \\
			\hline
			$(v/v_0)_\text{ini}$ & $1$ & $1$ & $1.0040 \pm 0.0065$ & $1.0095^{+0.0077}_{-0.0087}$ \\
			$\omega_a$ & $0$ & $< 0.000860$ & $0$ & $< 0.00177$ \\
			\hline
			$H_0$ & $67.61 \pm 0.43$ & $67.41 \pm 0.48$ & $68.2 \pm 1.1$ & $68.7 \pm 1.2$ \\
			$S_8$ & $0.825 \pm 0.012$ & $0.821 \pm 0.013$ & $0.826 \pm 0.013$ & $0.822 \pm 0.014$ \\
			$\sigma_8$ & $0.8096 \pm 0.0069$ & $0.8028^{+0.0094}_{-0.0082}$ & $0.815 \pm 0.011$ & $0.811 \pm 0.013$ \\
			$\phi_{\rm ini}$ & $0$ & $1.27^{+0.46}_{-0.72}$ & $0$ & $1.74^{+0.62}_{-1.1}$
		\end{tabular}
	\caption{Marginalized distributions of parameters in four models fitted with CMB$_\text{base}$+BAO. The uncertainties and the upper bounds of parameters are shown at 68\% and 95\% confidence levels, respectively. Note that $(v/v_0)_{\rm ini} = m_e/m_{e,0}$ in the $\Lambda$CDM+$m_e$ model.}
	\label{Tab:models_compare_nowl}
\end{table*}

\begin{table*}[htp]
	\begin{tabular}{c|c|c|c|c}
		Data & \multicolumn{4}{c}{CMB$_\text{base}$+BAO+WL}  \\
		\hline
		Model & $\Lambda$CDM & $\Lambda$CDM+$\omega_a$ & $\Lambda$CDM+$m_e$ & axi-Higgs  \\
		\hline\hline
		
		$\omega_b$ & $0.02248 \pm 0.00013$ & $0.02250 \pm 0.00013$ &  $0.02253 \pm 0.00016$ & $0.02270^{+0.00019}_{-0.00022}$\\
		$\omega_c$ & $0.11806 \pm 0.00088$  &  $0.11779 \pm 0.00091$ & $0.1186 \pm 0.0018$ & $0.1207^{+0.0022}_{-0.0027}$ \\
		$100\theta_\text{MC}$ & $1.04108 \pm 0.00028$ & $1.04111 \pm 0.00030$ & $1.0427 \pm 0.0046$ & $1.0505^{+0.0061}_{-0.0083}$ \\
		$\tau_\text{reio}$ & $0.0510 \pm 0.0075$ & $0.0529 \pm 0.0076$ & $0.0509 \pm 0.0076$ & $0.0527 \pm 0.0079$ \\
		$\ln (10^{10} A_s)$ & $3.031 \pm 0.015$ & $3.035 \pm 0.015$ & $3.032 \pm 0.015$ & $3.038 \pm 0.016$ \\
		$n_s$ & $0.9688 \pm 0.0036$ & $0.9692 \pm 0.0037$ & $0.9682 \pm 0.0042$ & $0.9672 \pm 0.0042$ \\
		\hline
		$(v/v_0)_\text{ini}$ & $1$ & $1$ & $1.0023 \pm 0.0066$ & $1.0135^{+0.0088}_{-0.012}$ \\
		$\omega_a$ & $0$ & $< 0.00133$ & $0$ & $< 0.00323$ \\
		\hline
		$H_0$ & $68.19 \pm 0.40$ & $67.80^{+0.53}_{-0.47}$ & $68.6 \pm 1.1$ & $69.4 \pm 1.3$ \\
		$S_8$ & $0.805 \pm 0.010$ & $0.800 \pm 0.011$ & $0.805 \pm 0.010$ & $0.797 \pm 0.012$ \\
		$\sigma_8$ & $0.8001 \pm 0.0062$ & $0.790^{+0.011}_{-0.0078}$ & $0.803 \pm 0.011$ & $0.794 \pm 0.013$ \\
		$\phi_{\rm ini}$ & $0$ & $1.68^{+0.67}_{-0.79}$ & $0$ & $2.6 \pm 1.1$
	\end{tabular}
	\caption{Same as Tab.~\ref{Tab:models_compare_nowl} but with CMB$_\text{base}$+BAO+WL.}
	\label{Tab:models_compare_base}
\end{table*}

\begin{table*}[htp]
	\begin{tabular}{c|c|c|c|c}
		
		Model & \multicolumn{2}{c|}{$\Lambda$CDM+$\omega_a$} & \multicolumn{2}{c}{$\Lambda$CDM+$m_e$} \\
		\hline
		Code & {\tt aHCAMB} & {\tt axionCAMB} & {\tt aHCAMB} & {\tt CAMB}+{\tt CosmoRec} \\
		\hline
		$\omega_b$ & $0.02220 \pm 0.00016$ & $--$ & $0.02256 \pm 0.00017$ & $0.02255 \pm 0.00018$ \\
		$\omega_c$ & $0.1198 \pm 0.0015$ & $0.119 \pm 0.002$ & $0.1208 \pm 0.0019$ & $0.1208 \pm 0.0018$ \\
		$m_e/m_{e,0}$ & $1$ & $--$ &  $1.0077 \pm 0.0069$ & $1.0078 \pm 0.0067$ \\
		$\omega_a$ & $< 0.00349$ & $< 0.003$ & $0$ & $0$ \\
		\hline
		$\tau_\text{reio}$ & $0.077 \pm 0.016$ & $--$ & $0.0552 \pm 0.0074$ & $0.0549 \pm 0.0074$ \\
		$\ln(10^{10} A_s)$ & $3.088 \pm 0.031$ & $--$ & $3.046 \pm 0.014$ & $3.045 \pm 0.014$ \\
		$n_s$ & $0.9645 \pm 0.0048$ & $--$ & $0.9656 \pm 0.0041$ & $0.9654 \pm 0.0040$ \\
		\hline
		$H_0$ & $65.9^{+1.6}_{-1.0}$ & $66.61 \pm 1.31$ &  $69.1 \pm 1.2$ & $69.1 \pm 1.2$ \\
		$\phi_{\rm ini} / M_{\rm pl}$ & $0.115 \pm 0.045$ & $0.11 \pm 0.04$ & $0$ & $0$
	\end{tabular}
	\caption{Marginalized posteriors of cosmological parameters reproduced from {\tt aHCAMB} compared with the previous implementations in $\Lambda$CDM+$\omega_a$ (see Tab.~3 of~\cite{Hlozek:2017zzf}) and $\Lambda$CDM+$m_e$ (see Tab.~1 of~\cite{Hart:2019dxi}) fitted with the data sets used in the corresponding works. {\tt aHCAM} independently yields consistent results with the other numerical codes.}
	\label{Tab:ahcamb_test}
\end{table*}

\begin{figure*}[b]
	\centering
	\includegraphics[scale=0.5]{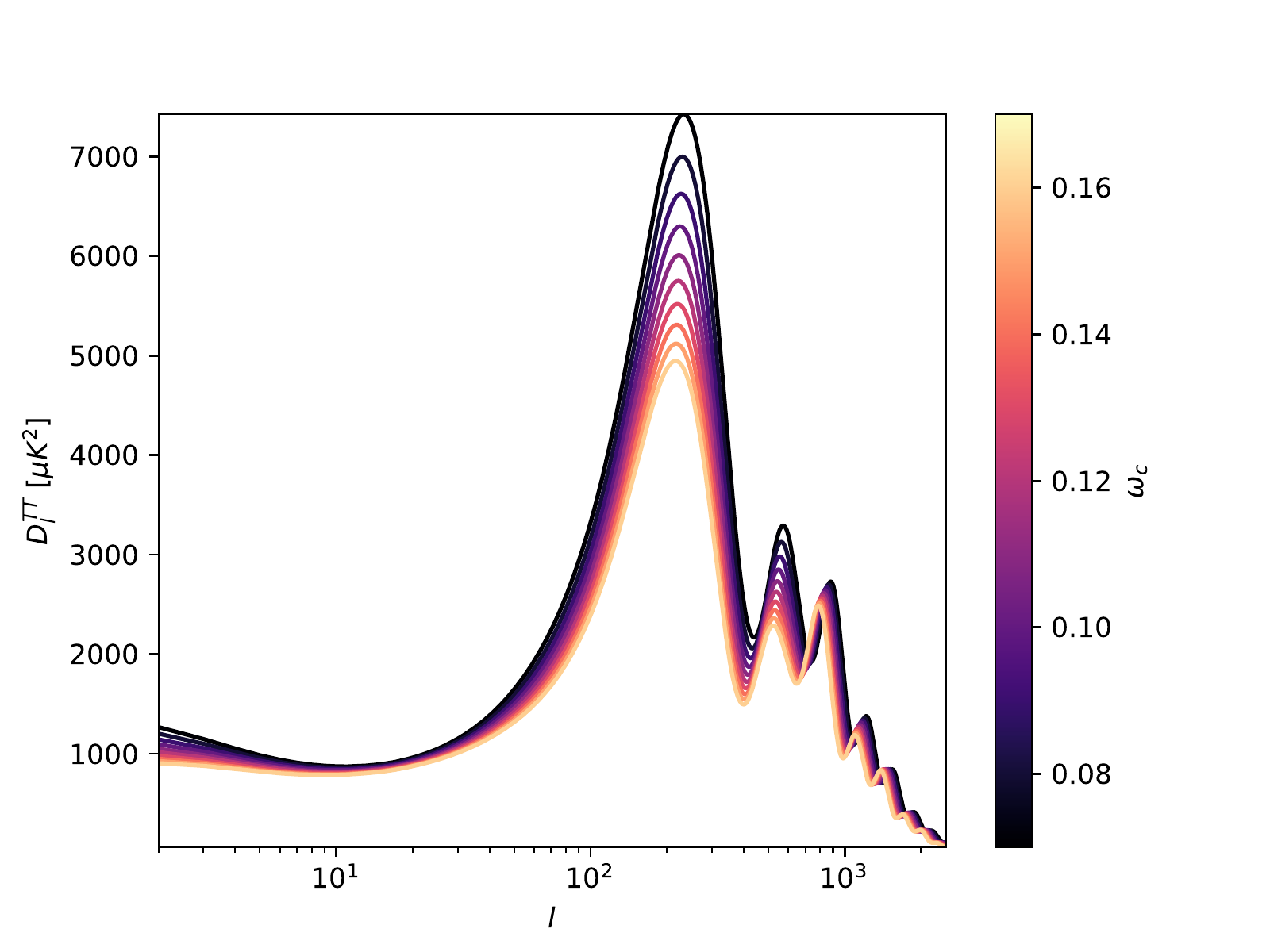}\includegraphics[scale=0.5]{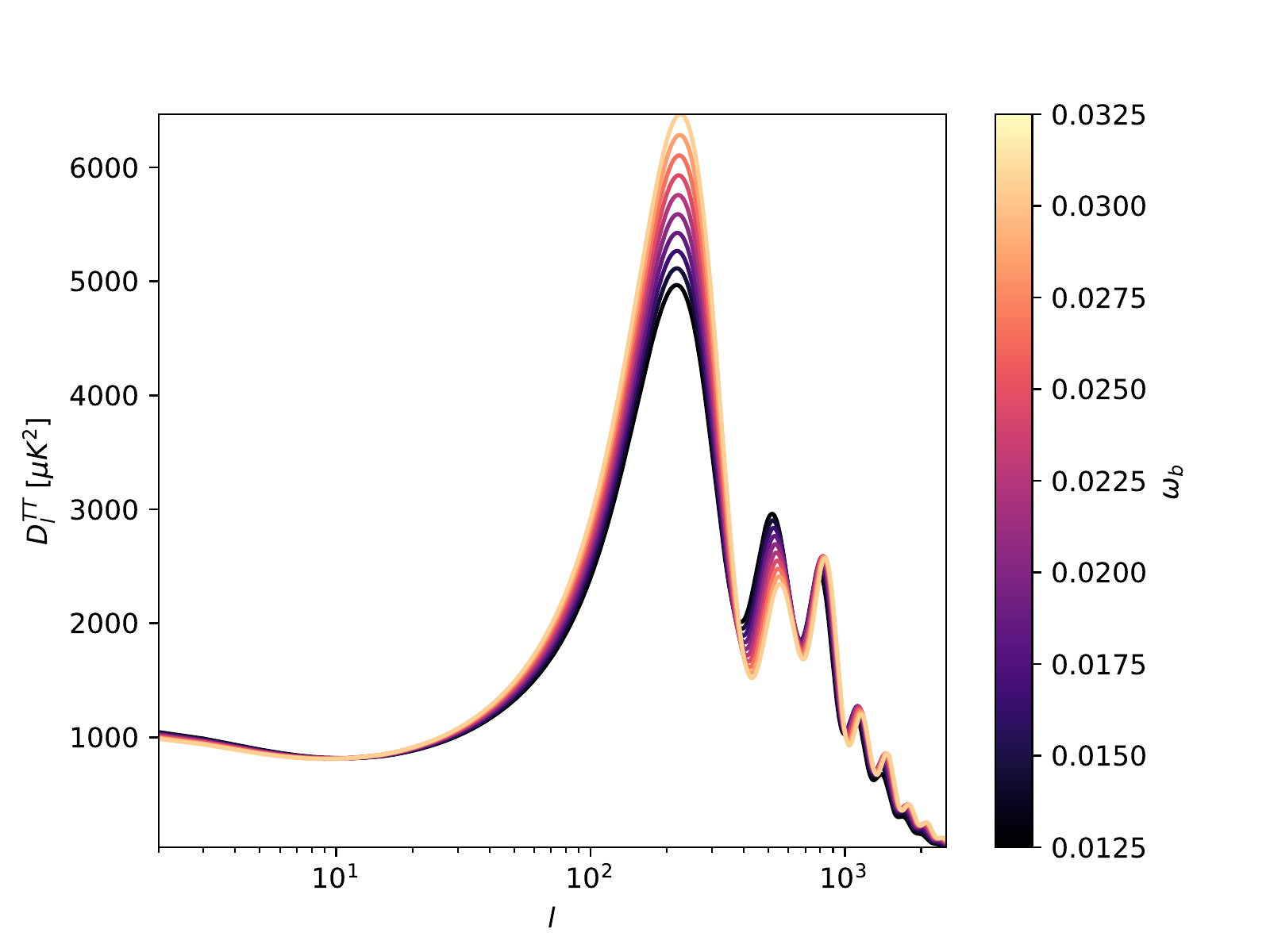}
	\includegraphics[scale=0.5]{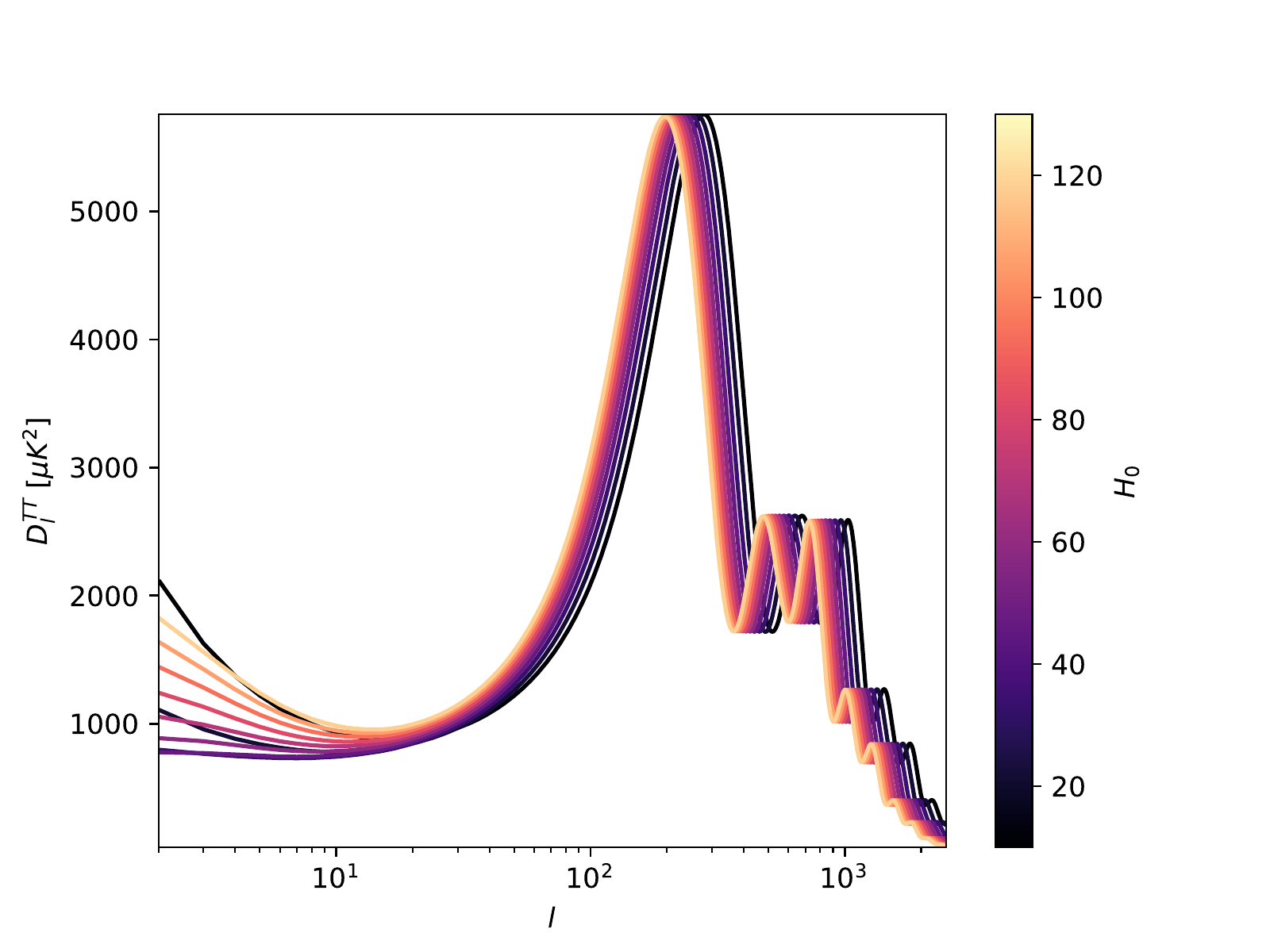}\includegraphics[scale=0.5]{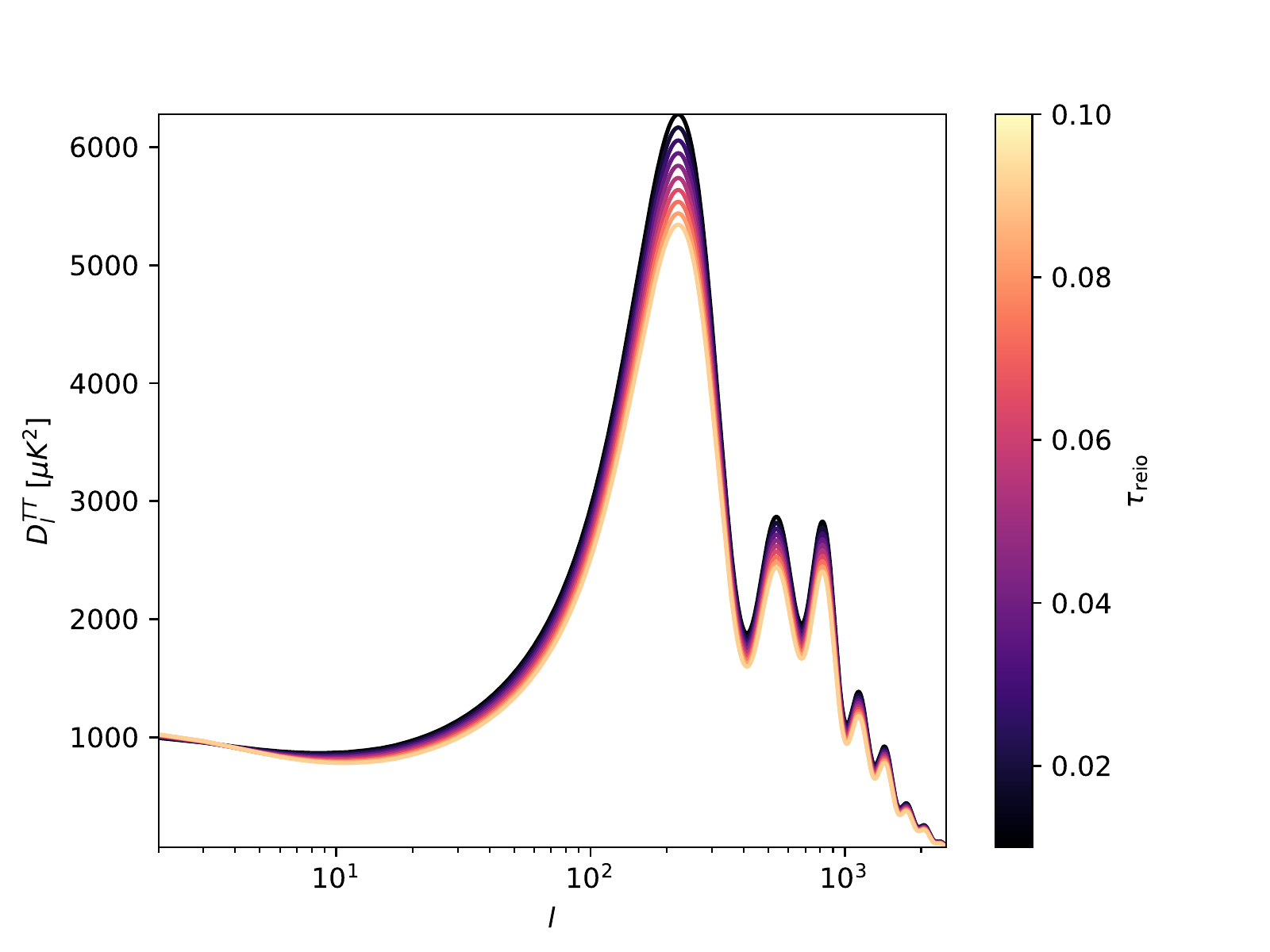}
	\includegraphics[scale=0.5]{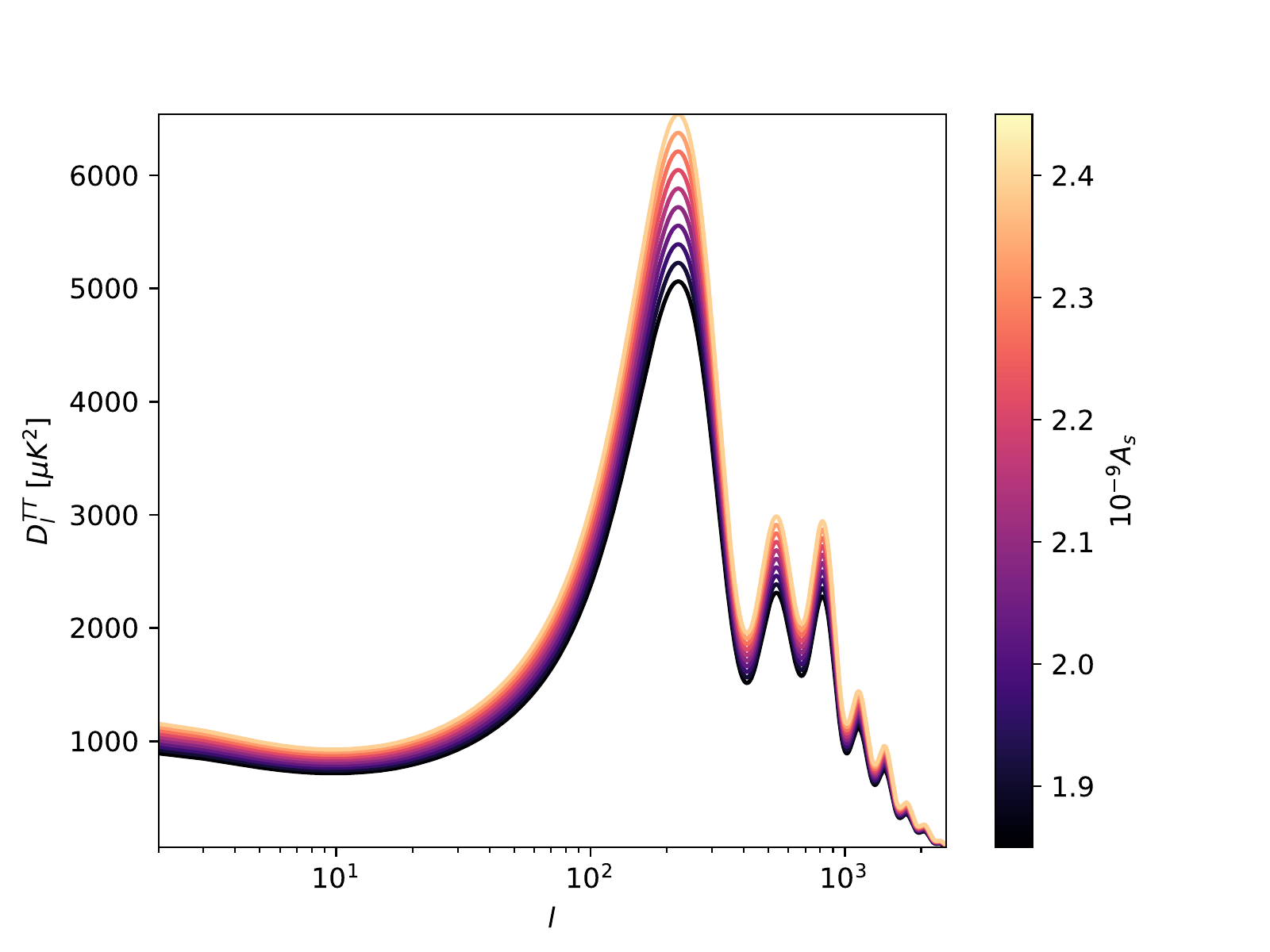}\includegraphics[scale=0.5]{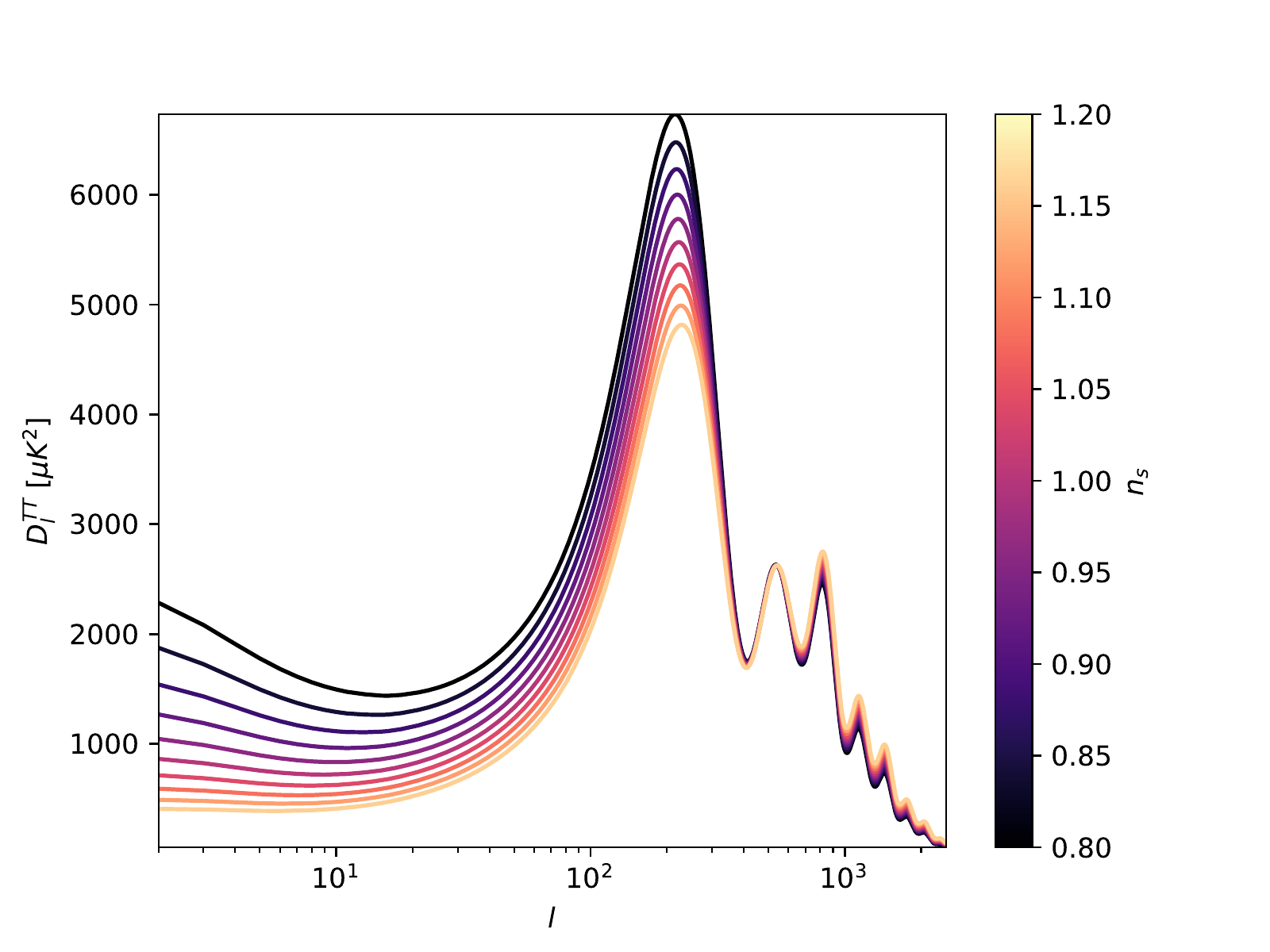}
	\caption{CMB temperature spectra variations with respect to six baseline parameters $\omega_b$, $\omega_c$, $H_0$, $\tau_{\rm reio}$, $A_s$, $n_s$ from top to bottom and left to right in respective order.}
	\label{Fig:cmb_spectra_extra}
\end{figure*}


\begin{figure*}[b]
	\centering
	\includegraphics[scale=0.5]{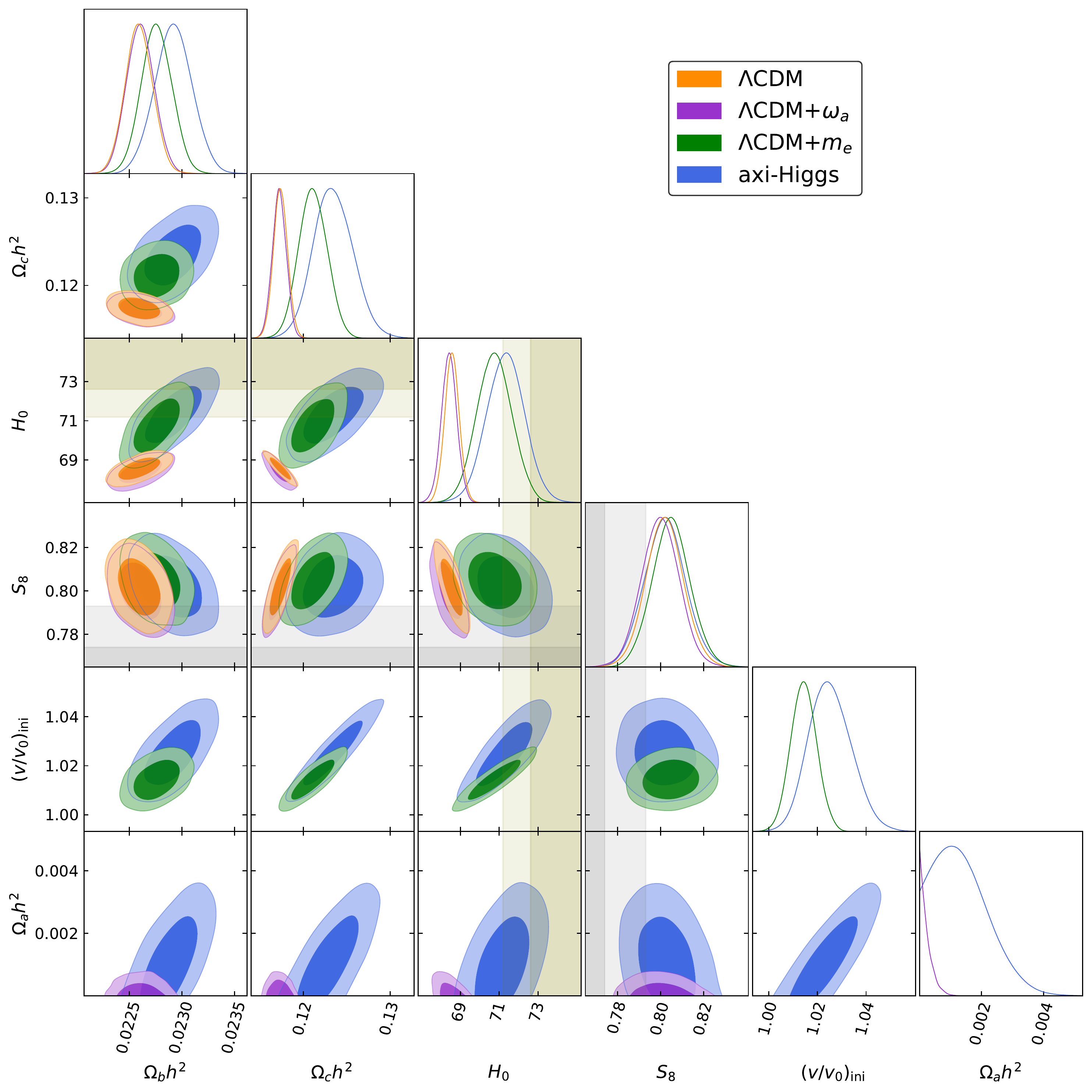}
	\caption{Posterior distributions of cosmological parameters for $\Lambda$CDM, $\Lambda$CDM+$\omega_a$, $\Lambda$CDM+$m_e$, axi-Higgs fitted with the combined data set of CMB$_\text{full}$+BAO+WL+SN+R19. The shaded olive and grey bands represent the local measurement of $H_0 = 74.03 \pm 1.42$ km/s/Mpc and the WL measurement of $S_8 = 0.755^{+0.019}_{-0.021}$, respectively.}
	\label{Fig:models_compare_full}
\end{figure*}


\bibliography{reference}

\end{document}